\documentclass[aps,amssymb,amsmath,reprint]{revtex4-1} 

\usepackage{graphicx}  
\usepackage{dcolumn}   
\usepackage{physics}
\usepackage{color}
\usepackage{appendix}
\usepackage[normalem]{ulem}

\definecolor{orange}{rgb}{0.8, 0.3, 0}

\definecolor{blueviolet}{rgb}{0.2, 0.2, 0.6}
\usepackage[pdftex, 
bookmarks=true, 
colorlinks=true,
allcolors=blueviolet,
pdfstartview={FitH}, 
]{hyperref}

\newcommand{\hethree}{$^3$\rm{He}}
\newcommand{\hefour}{$^4$\rm{He}}

\begin{document}
\title{Quantum bath suppression in a superconducting circuit by immersion cooling}
\author{M.~Lucas$^{1}$}
\author{A. V.~Danilov$^2$}
\author{L. V.~Levitin$^{1}$}
\author{A.~Jayaraman$^2$}
\author{A. J.~Casey$^{1}$}
\author{L. Faoro$^{3}$}
\author{A. Ya.~Tzalenchuk$^{1,4}$}
\author{S. E.~Kubatkin$^2$}
\author{J.~Saunders$^{1}$}
\author{S. E. de Graaf$^{4}$}
\email{sdg@npl.co.uk}

\affiliation{$^{1}$Physics Department, Royal Holloway University of London, Egham, United Kingdom } 
\affiliation{$^2$ Department of Microtechnology and Nanoscience MC2, Chalmers University of Technology, SE-412 96 G\"oteborg, Sweden}
\affiliation{$^3$Google Quantum AI, Google Research, Mountain View, CA, USA }
\affiliation{$^4$National Physical Laboratory, Teddington TW11 0LW, United Kingdom }

\begin{abstract}
\end{abstract}
\maketitle
{\bf{Quantum circuits interact with the environment via several temperature-dependent degrees of freedom. Yet, multiple experiments to-date have shown that most properties of superconducting devices appear to plateau out at $\bf T\approx 50$\,mK -- far above the refrigerator base temperature. 
This is for example reflected in the thermal state population of qubits \cite{PhysRevA.101.012336, PhysRevLett.114.240501,  PhysRevLett.124.240501, sultanov2021}, in excess numbers of quasiparticles \cite{PhysRevLett.121.157701}, and polarisation of surface spins \cite{degraaf2017} -- factors contributing to reduced coherence. We demonstrate how to remove this thermal constraint by operating a circuit immersed in liquid $^3$He. This allows to efficiently cool the decohering environment of a superconducting resonator, and we see a continuous change in measured physical quantities down to previously unexplored sub-mK temperatures. The $^3$He acts as a heat sink which increases the energy relaxation rate of the quantum bath coupled to the circuit a thousand times, yet the suppressed bath does not introduce additional circuit losses or noise. Such quantum bath suppression can reduce decoherence in quantum circuits and opens a route for both thermal and coherence management in quantum processors. 
}}

Thermal management is a central problem in computer engineering. This is true for classical processors, where inability to remove heat from transistors resulted in a stalled clock frequency for the last 20 years \cite{Denning}, and this is also true for superconducting quantum processors where various temperature-dependent factors limit their coherence. Scaling up quantum processors \cite{arute} inevitably exacerbates this problem and minimising the impact from all decoherence mechanisms at play is essential for achieving fault-tolerant quantum computing \cite{vepsalainen2020, mcewen2021}.

Cooling of devices operated in cryogenic vacuum represents a significant challenge because all solid-state cooling pathways -- through quasiparticles in the superconducting material and phonons both there and in the substrate -- become inefficient. A large body of experimental data indicates physical observables becoming temperature-independent below $\sim50$\,mK, well above the dilution refrigerator base temperature of $\sim10$\,mK. This is consistently seen in qubit state population \cite{PhysRevA.101.012336, PhysRevLett.114.240501, PhysRevLett.124.240501, sultanov2021}, qubit coherence times \cite{PhysRevLett.107.240501}, frequency flicker noise \cite{burnett2014, degraaf2018}, surface electron spin polarisation \cite{degraaf2017}, and qubit flux noise \cite{PhysRevLett.118.057702}.
Improvement may be achieved by reducing the heat load from various external sources, such as ionising radiation \cite{vepsalainen2020, cardani2021}, cosmic particles \cite{wilen2021, martinis2021}, and high frequency photons \cite{PhysRevLett.121.157701, barends2011, krinner2019}, by careful shielding and filtering. This approach has had a lot of success over the years and is still a subject of intense research and technical development. However, further progress cannot be achieved without taking due care of the circuit's material environment, for which, unexpectedly, further cooling can lead to increased noise and decoherence.

Although naively one would think that cooling a superconducting circuit to the lowest possible temperature would freeze out any noisy environment, this is only partly true. To suppress decoherence originating from equilibrium quasiparticles \cite{PhysRevLett.121.157701} or residual thermal qubit excitations \cite{PhysRevA.101.012336, PhysRevLett.114.240501, PhysRevLett.124.240501, sultanov2021} the temperature shall be significantly below relevant energy scales, i.e $T\ll 300$\,mK for a device operating at 7\,GHz. However, well below these temperatures other decoherence mechanisms, in particular that associated with the dielectric environment of the devices, come into play. 
Dielectrics contain defects, which act as two-level systems (TLS) and counter-intuitively, noise due to TLS increases upon cooling \cite{burnett2014, PhysRevB.91.014201}. 

\begin{figure*}
\centering
\includegraphics[width=1\textwidth]{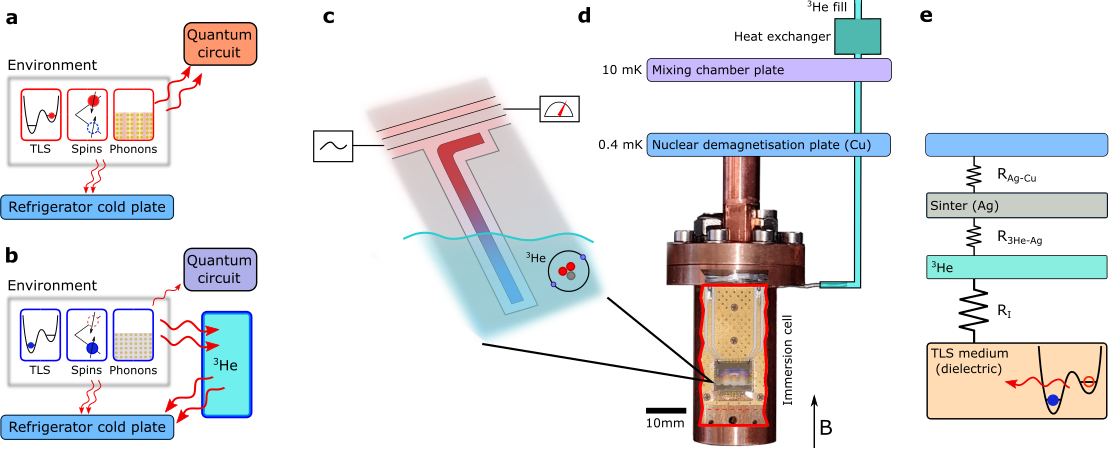}
\caption{\label{fig:fig1} {\bf{ Immersion of a superconducting quantum circuit in liquid $^3$He.}} a) In vacuum the environment of the quantum circuit is poorly thermalised to the cold plate of the refrigerator. b) When immersed in liquid \hethree, the cooling of the environment is significantly improved by $^3$He acting as a heat sink. c) A superconducting resonator, used in our measurements, taking the temperature of the decohering environment of quantum circuits. d) Experimental setup: The immersion cell containing the sample is thermally anchored to an adiabatic nuclear demagnetisation stage that reaches $T=400$\,\textmu K. The nuclear stage is mounted to the mixing chamber plate of a dry dilution refrigerator. e) Energy relaxation pathway from the TLS bath to the cold plate via \hethree~and silver sinter. The link between TLS medium and liquid \hethree~is the bottleneck for further quantum bath suppression.}
\end{figure*}

Here we present a radically different route to approach these challenges by immersion cooling of a superconducting circuit in liquid \hethree. We show that \hethree~provides an efficient heat sink for the circuit environment and dramatically increases the energy relaxation rate of the TLS bath, while otherwise appearing essentially inert to the quantum circuit itself. This opens up multiple ways in which significant improvement in circuit coherence may be achieved, both by cooling and by suppressing coherence in the noisy environment.
Future optimisation of such quantum bath suppression using \hethree~may lead to significantly reduced noise also at dilution refrigerator base temperatures.

Experimentally, our approach is to use planar superconducting resonators, which have emerged as a convenient platform to interrogate the decohering environment \cite{degraaf2017, burnett2014,brehm2017, bejanin2022, PhysRevApplied.17.034025, degraaf2021,PhysRevMaterials.1.012601, degraaf2022}. 
In particular, the amplitude of the low-frequency $1/f$ frequency noise is very sensitive to the TLS temperature \cite{PhysRevB.91.014201}. Additionally, the temperature of the surrounding spin bath reveals itself in the electron spin resonance (ESR) spectrum measured via field-dependent losses of the same resonators.
When the resonator is immersed in \hethree, we observe improved thermalisation of the TLS in the noise measurements and of the spin bath in the ESR measurements, as illustrated in Figure \ref{fig:fig1}.

Derived from recent advances in ultra-low temperature technology and the cooling of electronic systems to sub-mK temperatures \cite{jones2020, levitin2022} we construct an immersion cell suitable for a superconducting quantum circuit.
Cooling is achieved by placing the circuit, in our case a NbN superconducting resonator \cite{PhysRevApplied.14.044040} on a sapphire substrate, inside the immersion cell, as shown in Figure \ref{fig:fig1}d. The superfluid leak-tight copper cell with RF feedthroughs and extensive RF filtering provides a well controlled microwave environment. It is thermally anchored to the experimental plate of an adiabatic nuclear demagnetisation refrigeration (ANDR) stage attached via a superconducting heat switch to the lowest temperature plate (10 mK) of a dry dilution refrigerator \cite{nyeki2022}. The experimental plate of the ANDR can reach  temperatures of $\approx 400$\,\textmu K, as measured using SQUID noise thermometry \cite{nyeki2022}. The cell can be filled with \hethree~via a thin capillary. To ensure good thermalisation of the liquid \hethree~to the cell's metal enclosure silver sinter heat exchangers are implemented (See SI for further details). For ESR spectroscopy experiments a magnetic field (B) up to 0.5 T parallel to the sample surface could be applied. We refer to Supplemental Information (SI) for details on our measurement techniques.

Reliable thermometry is an essential pre-requisite for interpretation of ultra-low temperature data. On-chip ESR not only reveals the presence of unwanted surface spins coupling to the resonator through their magnetic moments (a source of flux noise \cite{PhysRevLett.110.147002, PhysRevLett.118.057702}), but also serve as an intrinsic thermometer in the relevant temperature range.
\begin{figure}
\centering
\includegraphics[width=0.4\textwidth]{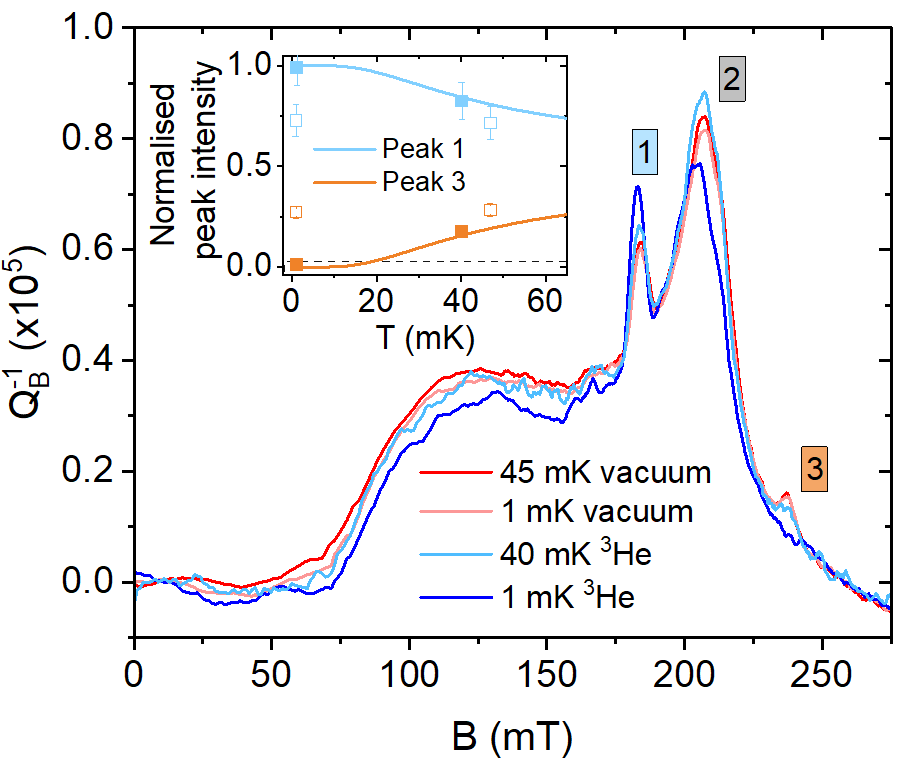}
\caption{\label{fig:esr} {\bf{Cooling of surface electron spins.}} Continuous wave electron spin resonance spectra of surface spins intrinsic to a 5.85 GHz resonator, measured with an average number of photons circulating in the resonator of $\langle N\rangle\approx 200$. Inset: Normalised intensity of the hyperfine-split atomic hydrogen peaks (labelled 1 and 3) versus nuclear stage noise thermometer temperature. Empty symbols represent measurement in vacuum and filled symbols in \hethree. Error bars are propagated errors from fitting the peak intensities. Solid lines are the expected peak intensities based on the thermal population of ESR levels hyperfine-split by $A=1.42$ GHz. The dashed line is an estimate of the minimum sensitivity of our technique, below which we could not detect the third peak.}
\end{figure}
To this end we show in Figure \ref{fig:esr} that, unlike previous experiments on spins coupled to quantum circuits where the spin polarisation was saturated at about $T=50$ mK \cite{degraaf2017}, surface spins are cooled to much lower temperatures in the presence of \hethree, with no other apparent change in the ESR spectra. 
The measured ESR spectrum is rather complex, consisting of many different species, and has been discussed in detail previously \cite{degraaf2017, un2022}. Here we focus on the species that are most suitable for intrinsic thermometry at these low temperatures, namely the two peaks labelled 1 and 3 that arise from atomic hydrogen \cite{degraaf2017}.  The hyperfine interaction in the hydrogen atom results in two electronic spin transitions separated in energy by $1.42$ GHz ($=68$~mK), with relative intensity that follows the Boltzmann distribution. Thus if spins are cooled to zero temperature the transition involving the higher energy level transition (peak 3) will vanish, the trend clearly seen in figure \ref{fig:esr} in the presence of \hethree. 
Having established improved thermalisation of surface spins we now turn to the TLS bath that couples through charge dipoles to the same circuit.

\begin{figure*}
\centering
\includegraphics[width=0.83\textwidth]{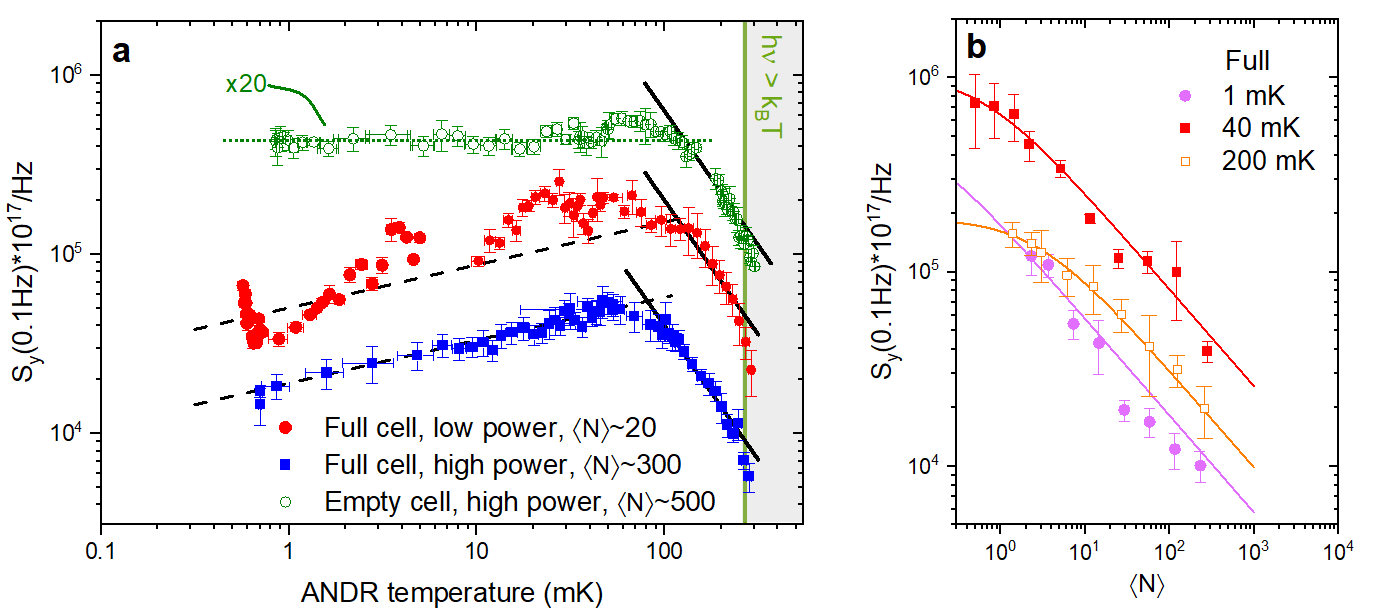}
\caption{\label{fig:noise} {\bf{Cooling the TLS bath by immersion into $^3$He reduces noise.}} a) The magnitude of the $1/f$ frequency noise power spectral density $S_y(f)$ of a $\nu_0=6.45$ GHz superconducting resonator evaluated at $f=0.1$ Hz versus nuclear stage temperature for two selected microwave drive powers (average photon number $\langle N\rangle$) with the cell full of \hethree~(filled markers) and empty cell (empty markers). The latter has been scaled by a factor 20 for better visualisation (see SI for unscaled version). Each dataset is a single temperature ramp taking $\approx 6$ days. Solid and dashed slopes show $T^\beta$ in the low and high temperature regimes respectively, with $\beta=0.25$ and $-1.5$ respectively. Horizontal dashed line is a guide for the eye. 
b) Photon number dependence of the noise with \hethree~at selected temperatures taken across shorter timescales (5 hours per temperature). Solid lines are fits to the expected dependence of the noise ($\propto(1+\langle N\rangle/N_c)^{-1/2}$) where the weak fields regime with a levelling-off to a constant noise versus $\langle N\rangle$ is evident at high temperature. Full noise spectra across all timescales can be found in SI.}
\end{figure*}
Figure \ref{fig:noise}a compares the temperature dependence of the $1/f$ frequency noise of a 6.45 GHz resonator with vacuum or \hethree~in the sample cell (for more data on different devices, see SI).
Similar to many previous experiments \cite{burnett2014, burnett2016,  degraaf2018, ramanayaka, degraaf2020}, in vacuum the noise increases on cooling according to a power law  $~T^{-1.5}$ followed by saturation to a constant level below $\sim 80$ mK due to insufficient thermalisation.

When the cell is filled with \hethree~the situation is very different. The noise changes with fridge temperature all the way down to 1 mK. Above 100~mK the magnitude and temperature dependence of the noise is the same in vacuum and in \hethree, but below a certain crossover temperature $T_{x}\sim 80$ mK the noise instead starts to decrease with reduced temperature according to a power law $T^{0.25}$.   
Remarkably, \hethree~immersion appears to break the predicted \cite{PhysRevB.91.014201} trend of increasing noise with cooling (otherwise expected to persist to well below 10 \textmu K, see below). The noise measured at 1~mK is more than three orders of magnitude below this expected $T^{-1.5}$ trend. 

\begin{figure*}
\centering
\includegraphics[width=0.85\textwidth]{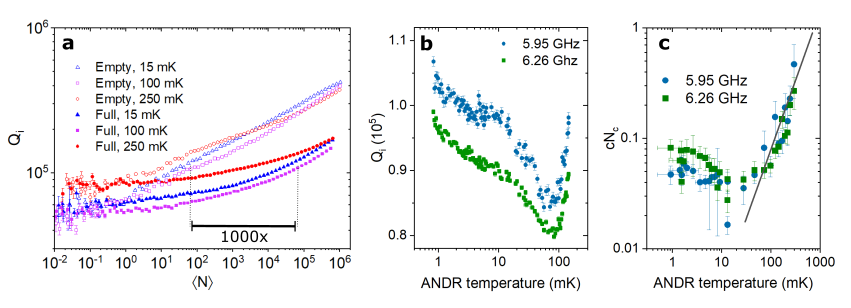}
\caption{\label{fig:q} {\bf{$^3$He increases TLS relaxation.}} a) Comparison of the TLS-limited internal quality factor for a 6.26 GHz resonator with and without \hethree~in the cell, for three temperatures. \hethree~increases the power needed to saturate to a given $Q_i$ by a factor $\sim 1000$. 
 b) The change in internal Q vs temperature for a fixed drive power of $\langle N\rangle\sim 10^4$. c) Extracted critical photon number $N_c$ times a prefactor $c$, from fitting the $Q_i(N)$ data to $1/Q_i \propto {\rm{ln}}(cN_c/\langle N \rangle)$. Solid line shows $T^{1.25}$, the expected scaling of $\Gamma_2$.}
\end{figure*}

A further striking effect of immersing the circuit in \hethree~is revealed in the dependence of the internal quality factor $Q_i$ of the resonators on the microwave power (photon number, $\langle N\rangle$), presented in Figure \ref{fig:q}a for three temperatures. Noticeably, \hethree~does not affect $Q_i$ at the single photon level, meaning that the number of TLS present and their coupling to the resonator remains unchanged. Both for resonators in vacuum and in \hethree~the microwave excitation power increases $Q_i$ -- a well-known effect of TLS saturation -- but for resonators immersed in \hethree~the same $Q_i$ is achieved with $\sim1000$ times higher power; i.e. we find a dramatic increase in the characteristic TLS saturation power by three orders of magnitude.  
Figure \ref{fig:q}b showing the $Q_i$ extracted at a fixed drive power in the saturated regime indicates that there is a weak but steady dependence (and hence cooling) down to $<1$ mK. Furthermore, we also here observe a crossover occurring around $T_x\sim 80$\,mK. 

A notorious challenge in noise measurements (and more generally in operating quantum circuits requiring long-term stability \cite{kilmov2018, burnett2019, schlor2019, arute}) is inherent instabilities of TLS energies on longer timescales (spectral drift). 
This is particularly evident in the low power data in Figure \ref{fig:noise}a. To circumvent this problem, we measure noise at somewhat higher photon number which saturates the most strongly coupled fluctuators \cite{niepce2021}. Yet, we stay in a moderately weak fields regime \cite{PhysRevB.91.014201}, as evidenced by the power dependence of the noise shown in Figure \ref{fig:noise}b.
In the low power data in Figure \ref{fig:noise}a the measured noise level varies on top of the general trend by a factor 2-3 during the course of the measurement, which takes $\sim 6$ days; however, the overall trend remains unchanged.

To understand the full body of experimental data we first focus on the region 100-250~mK where the noise is well understood. Here the noise is increasing upon cooling both in vacuum and in \hethree, consistent with previous observations and fully captured by the generalised tunneling model (GTM) \cite{PhysRevB.91.014201} for interacting TLS defects.

Here both energy loss and $1/f$ noise arise from the resonator coupling to a large number of coherent (near-)resonant TLS defects. These TLS drain energy from the resonator and dissipate it to substrate phonons, a process that determines $Q_i$, a measure of the average energy loss into the whole TLS bath.
They also, through their coherent coupling, mediate frequency fluctuations from the environment: the resonant coherent TLS are subjected to thermally activated spectral diffusion due to the interaction with a bath of incoherent, low energy ($E\ll k_BT$) TLS ("thermal two-level fluctuators", TLF) that incoherently flip-flop between two states, giving rise to both noise and the coherent TLS width $\Gamma_2$.  
Strongly coupled TLS contribute more strongly to the noise and they are also more easily saturated by microwave fields. In the weak microwave field regime (small $\langle N\rangle$) the magnitude of the noise is governed by the TLS dephasing rate $\Gamma_2\propto T^{1+\mu}$ arising from the coupling to the TLF bath (and independent of the TLS relaxation rate $\Gamma_1$), which yields a temperature dependence $S_y\propto T^{-1-2\mu}$ \cite{degraaf2018, PhysRevB.91.014201} (for $k_BT<h\nu_0$). Here $\mu$ is a small positive number characterising the nonlinear density of states of TLS arising from their interactions. From the data in Figure \ref{fig:noise} we find $\mu=0.25$, consistent with previous experiments \cite{burnett2014, degraaf2018, burnett2016}. Since the magnitude and the temperature dependence of the noise is the same in vacuum and in \hethree~above 100 mK we conclude that \hethree~does not influence $\Gamma_2$.

This leads to the remarkable conclusion that the enormous change in saturation power observed, Figure \ref{fig:q}a, means that \hethree~increases the average TLS relaxation rate $\Gamma_1$ $\sim 1000$ times because the critical number of photons for saturation of the TLS bath scales as $N_c  \propto\Gamma_1\Gamma_2$.

We now turn to the regime at low temperatures, below  $T_{x}\sim 80$~mK. 
First, we consider the implications of a significantly increased $\Gamma_1$ of the TLS bath.
In dielectrics the TLS excitation and relaxation occurs via interaction with phonons which couple via strain field. The relaxation rate can be expressed using the Golden rule formula as $\Gamma_1^{ph}=({M^2\Delta_0^2E})/({2\pi\rho\hbar^4v^5})\times\coth{\frac{E}{2k_BT}}$ \cite{phillips},
where $M$ is the deformation potential, $\Delta_0$ is the TLS tunneling matrix element, $E$ is the TLS energy, $\rho$ is the density and $v$ is the speed of sound of the material. For a resonator in vacuum, the dissipation is through the emission of phonons into the dielectrics hosting the TLS, and at the relevant temperatures this process is temperature independent with a rate that can be estimated to $\Gamma_1^{ph}\approx 10^2-10^3$ Hz \cite{PhysRevB.91.014201}. This is much smaller than the TLS dephasing rate due to interactions $\Gamma_2$: Previous estimates \cite{degraaf2018, burnett2016} yielded $\Gamma_2\approx 10^6-10^7 $ Hz at $T=50-100$ mK in similar devices.

Assuming the $\Gamma_2 \propto T^{1+\mu}$ dependence persists to lower temperatures means that in vacuum we would reach the regime of relaxation limited dephasing, $\Gamma_2\simeq 2\Gamma_1$, of the TLS bath below $10~\mu$K, which is experimentally inaccessible. \hethree~immersion increases the average $\Gamma_1$ to $\sim 10^5-10^6$ Hz, which increases this crossover temperature to $10-100$\,mK. 
This agrees with the observed crossover temperature in the noise and in the dependence of the quality factor on power $Q(\langle N\rangle)$. Furthermore, within the GTM $1/Q_i \propto {\rm{ln}}(cN_c/\langle N \rangle)$ \cite{PhysRevLett.109.157005}, where $c$ is a constant. The $Q(\langle N\rangle)$ data fits remarkably well to this logarithmic power dependence (see SI). In Figure \ref{fig:q}c we show that the temperature dependence of $cN_c$ follows the predicted $N_c\propto \Gamma_1\Gamma_2(T)\sim T^{1+\mu}$ scaling at high temperatures. However, below $\sim 80$ mK, around $T_x$, this trend changes abruptly, and becomes temperature independent. In the relaxation limited regime the noise is not expected to increase upon cooling, yet a temperature dependence may be inherited from mechanisms contributing to $\Gamma_1$, such as the \hethree-TLS interaction. \hethree~immersion thus prevents the TLS noise from rising more than three orders of magnitude upon cooling to 1 mK. 

A second scenario that in addition may account for the apparent reduction in the noise is TLS saturation. Such situation could arise because the measurement is conducted at a fixed driving power. As $\Gamma_2$ (and hence $N_c$) becomes smaller at lower temperatures the applied power more easily saturates the TLS because they become more coherent \cite{burin2015, PhysRevB.91.014201}. The GTM predicts a universal $T^{(1-\mu)/2}=T^{0.375}$ scaling of the noise in this regime, and power broadening would also result in the observed crossover in $N_c$ from $T^{1+\mu}$ to constant in temperature \cite{PhysRevLett.109.157005} (Figure \ref{fig:q}c) as for the relaxation limited scenario. Because we have significantly increased the average $\Gamma_1$ of TLS in the bath, one would think that this scenario is of less relevance in \hethree. 
Indeed, another important observation is that for saturation in the regime  $\Gamma_1\ll \Gamma_2$ the crossover temperature $T_x$ should depend on driving power, contrary to our data.

Yet, in any practical device the spatial variations in electric fields, the distribution of TLS parameters, and the fact that not all TLS are located in proximity to the exposed surface where they can couple to \hethree~means there still will exist TLS that are not suppressed by \hethree~and are therefore easily saturated. This prompts device improvements where surfaces and edges with strong electric fields should be placed in proximity to \hethree. 

As a first step to understand the \hethree-TLS interaction we note the long-standing problem of the thermal boundary resistance between solids and helium liquids, where the details of the interface, such as surface roughness \cite{ramiere2016} and the nature of the surface boundary layer, including the presence of 1-2 layers of solid helium at the interface due to van der Waals attraction \cite{scholtz,birchenko}, play a key role \cite{nakayama}.
Perhaps more closely related to this work are earlier acoustic and thermal measurements on strongly disordered  \cite{schubert} and porous \cite{PhysRevLett.52.1790, cheng_2013} materials immersed in helium that also found evidence of faster TLS relaxation. It has been suggested \cite{kinder} that one mechanism by which phonons in helium couple to TLS is via van der Waals interaction. The upper bound for the relevant deformation potential  in \hefour~was deduced to be $M\lesssim 2$ meV \cite{schubert} compared to $\approx 1$ eV for phonons in a solid.
Using these numbers we can attempt to roughly estimate the enhanced TLS relaxation rate in \hethree, compared to the sapphire substrate.
For sapphire we use $\rho = 4\times 10^3$ kg/m$^3$, $v = 1\times 10^4$ m/s, $M = 1$ eV. Similar values are also expected for TLS in the NbN surface oxide. For \hethree~we use $\rho = 60$ kg/m$^3$, $v= 200$ m/s and $M=1$ meV \cite{cheng_2013}. This yields $\Gamma_1^{^3He}/\Gamma^{sap}_1 \approx 10^4$ -- an order of magnitude larger than experimentally observed. This is not very surprising given the crudeness of the estimates and the fact that we measure the average for the whole TLS bath.  
Moreover, we note that below $\sim100$ mK the propagating acoustic modes in \hethree~are that of zero sound \cite{landau}. Zero sound modes and the nuclear magnetism \cite{PhysRevB.41.11011, PhysRevLett.31.76, PhysRevB.54.R9639} of \hethree~offers various interaction mechanisms with relevant degrees of freedom and a much richer spectrum of low energy excitations than in \hefour~\cite{dobbs_book}. To the best of our knowledge, the TLS-\hethree~coupling has not been studied in detail before, and at low temperatures other types of interactions may become as important as phonons, such as direct interaction between surface TLS and quasiparticles in \hethree~\cite{PhysRevLett.41.1487, kinder}. 

Understanding the mechanism at play is crucial for future improvements, and two further experiments (details in SI) suggests that phonon relaxation into \hethree~following the Golden rule alone does not capture the full picture. 
i)  Measurements with only a thin ($\sim4$\,nm) film of \hethree~covering the sample allow us to separate the two roles played by \hethree, namely to enhance TLS relaxation and to mediate cooling. For a thin \hethree~film we still observe the big change in saturation power (\hethree-TLS interaction) but a plateaued noise as in vacuum, indicating poor thermalisation. 
ii) Increasing the pressure of the \hethree~to 5 bar, whereupon both $\rho$ and $v$ increase by $\sim30$\% compared to standard vapour pressure \cite{dobbs_book}, should result in an almost five-fold reduction of $\Gamma_1$. Contrary, we observed a very moderate {\it increase} in saturation power ($<20\%$).

 Finally we turn to the dielectric properties of \hethree~to understand its compatibility with state of the art qubit circuits. The resonator frequency shift due to filling the cell agrees with the \hethree~dielectric constant  $\varepsilon_{r}=1.0426$ \cite{saitoh2004} within 1 part in 1000 (see SI).
 Liquid \hefour~has a low-temperature dielectric loss tangent $\tan\delta <5\times 10^{-6}$ at 9 GHz \cite{smorodin2016}. Similar values are expected for \hethree, however, to the best of our knowledge the this value not been reported at GHz frequencies.
 From the change in single-photon $Q_i$ at 10 mK as the cell is filled with \hethree~we estimate an upper bound for the loss tangent of $\tan\delta \ll 1.5\times 10^{-5}$ at 5.8 GHz, comparable to the best substrate dielectrics used.  
Likely $\tan\delta$ is much lower as significant TLS-induced parameter drift occurs between measurements, the main source of error in our estimate. The bound on the loss tangent translates to a limit for qubit coherence times of $T_1\gg 110$ $\mu$s for a 6~GHz qubit, i.e \hethree~is compatible with state of the art quantum circuits.

In conclusion we have shown that \hethree~is an efficient, low-loss cooling medium for quantum circuits and can cool down environmental degrees of freedom of the circuit: namely surface spins and the TLS bath. We also discovered the crucial role of \hethree~in suppressing the coherence of the TLS bath while otherwise being essentially inert to the circuit itself. Understanding details of the mechanisms at play will require further theoretical and experimental work. The rich phase diagram of \hethree~provides an exciting playground for bath engineering of quantum circuits, with multiple {\it in situ} tuning parameters to unpick the underlying physical mechanisms.
\hethree~immersion thus opens up a new avenue for exploring the origins of decoherence in quantum circuits and a promising pathway to further suppressing it. 

\let\oldaddcontentsline\addcontentsline
\renewcommand{\addcontentsline}[3]{}
\section*{Acknowledgements}
 We thank X. Rojas for his help with the design of the microwave setup. This work was supported by the UK government department for Business, Energy and Industrial Strategy through the UK national quantum technologies programme,  
The Swedish Research Council (VR) (grant agreements 2016-04828,  2019-05480 and 2020-04393) and Knut and Alice Wallenberg Foundation via the Wallenberg center for Quantum Technology (WACQT). 
The research leading to these results has also received funding from the European Union’s Horizon 2020 Research and Innovation Programme, 
under Grant Agreements No. 824109 and No. 766714/HiTIMe. S.D.G. acknowledges support by the Engineering and Physical Sciences Research Council (EPSRC) (Grant Number EP/W027526/1).

\bibliography{main}
\let\addcontentsline\oldaddcontentsline
\clearpage

\renewcommand{\theequation}{S\arabic{equation}}
\renewcommand{\thefigure}{S\arabic{figure}}
\renewcommand{\thetable}{S\arabic{table}}
\begin{centering}{\large{\bf{SUPPLEMENTARY INFORMATION}\\}}\end{centering}\vspace{3mm}

Experiments on quantum circuits at microkelvin temperatures is a cross-disciplinary effort. In this Supplementary we provide a detailed account of the cryogenic, microwave, metrology, and analytical background of our research in order to catalyze future development in this direction.

{\let\clearpage\relax \tableofcontents}

\section{Experimental setup}
\subsection{\label{sec:the dilution refrigerator}Refrigerator and thermometry}
The experiment was conducted on a cryogen-free Triton 200 Oxford Instruments dilution refrigerator (DR). This DR has been fitted with an adiabatic nuclear demagnetisation stage that allows it to operate at temperatures as low as $400\,$\textmu K. This system will be referred to as ND4 and is described in detail in \cite{nyeki2022}.

The experiment is installed in the ultra-low RF-noise environment of ND4, on the adiabatic nuclear demagnetisation refrigerator plate (ANDRP), allowing it to be cooled below 1\,mK.

In addition to the RuO\textsubscript{2} thermometer on the mixing chamber plate (MCP), ND4 is also equipped with two current sensing noise thermometers (CSNT) with SQUID readout: one on the MCP and one on the ANDRP. CSNT allows to accurately measure temperatures down to $100\,$\textmu K\cite{lusher2001,shibahara2016}. We used a low resistance sensor ($\approx2\,\rm m\Omega$) for the ANDRP CSNT to reduce the error on the measured temperature originating from the background noise of the SQUID. The temperature is obtained by fitting the power spectral density averaged over fifty noise traces. Each noise trace is sampled $2^{20}$ times at $200\times10^3\,$samples/s, which gives an acquisition time of $5.24\,$s per trace and a temperature measurement every $262\,$s. This brings the error in temperature measurement below $1\,$\% at $0.4\,$mK and below $0.1\,$\% at $10\,$mK.

\subsection{\label{sec:the microwave system}Microwave wiring}
\begin{figure}
\includegraphics[width=0.9\linewidth]{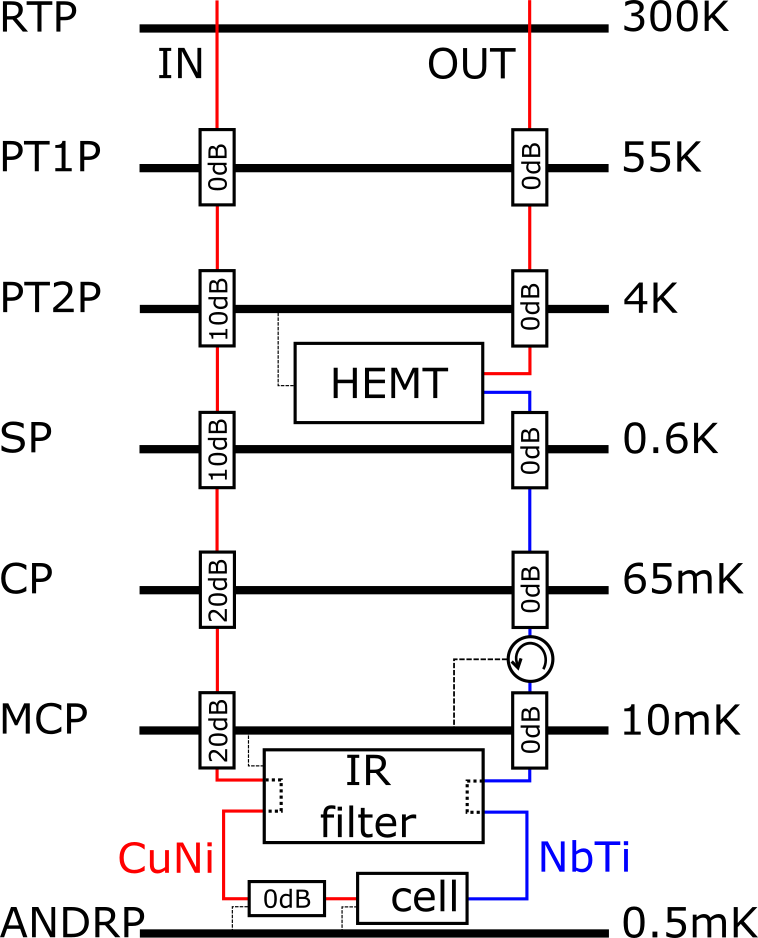}
\caption{\label{fig:rf}Microwave installation used for the present experiment. 'RTP': room temperature plate, 'PT1P': pulse tube stage 1 plate, 'PT2P': pulse tube stage 2 plate, 'CP': cold plate (or $100\,$mK plate), 'MCP': mixing chamber plate, 'ANDRP': adiabatic nuclear demagnetisation refrigerator plate. Red(blue) lines represent the CuNi-CuNi(NbTi-NbTi) UT-034 coax lines. The smaller rectangular boxes show the configuration of attenuators. 'IR filter' is a eccosorb infra-red filter. The circle with the circular arrow between MCP and CP is a triple-junction isolator. 'HEMT' is a high electron mobility transistor amplifier. The dotted lines represent the thermalisation link to the different temperature stages.}
\end{figure}

The microwave installation in the cryostat used for the measurements was built following the recipe detailed in \cite{krinner2019}, extended for  operation compatible with temperatures below $1\,$mK. Figure~\ref{fig:rf} shows an overview of the whole microwave assembly, which consists of one input line and one output line. UT-034 ($0.86\,$mm OD) NbTi-NbTi semi-rigid superconducting cryogenic coax cables are used for the connections between the cell and the HEMT amplifier on the $4\,$K plate, and UT-034 CuNi-CuNi semi-rigid cryogenic coax cables elsewhere.

Both lines are thermalised to each temperature stage with attenuators (the choice of attenuation will be discussed later). Infra-red (IR) frequencies are filtered out from both lines with an infra-red eccosorb filter, which is thermalised to the MCP.
The output signal is amplified with a cryogenic HEMT amplifier (LNF-LNC4\_8C) with a gain at 4\,K of 41\,dB and a noise temperature at 6\,GHz of 1.5\,K, installed on the pulse tube stage 2 plate (PT2P) at $4\,$K.
To prevent thermal radiation from the HEMT reaching the experiment without attenuating the outgoing signal, a triple junction isolator (LNF-ISISISC4\_8A) with $\text{S}_{21}(6\,\text{GHz}, 3\,\text{K})=-0.1\,$dB and an isolation at $(6\,\text{GHz}, 3\,\text{K})\ge70\,$dB is placed between the MCP and the still plate (SP). 

\begin{figure}
\includegraphics[width=1\linewidth]{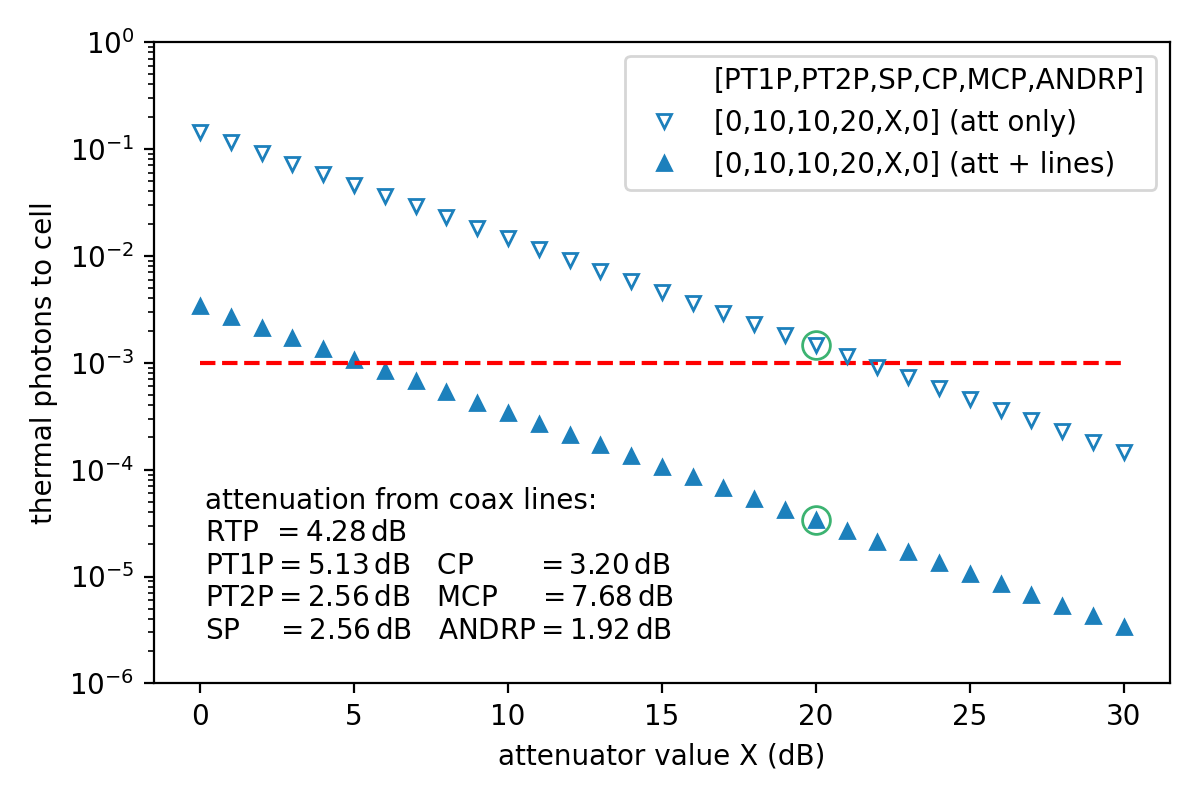}
\caption{\label{fig:attenuation}The population of thermal photons at $6\,$GHz reaching the cell for different values of the input line MCP attenuator. The empty markers show the population of thermal photons calculated without taking into account of the attenuation in the lines, as an upper bound. The solid markers show the population of thermal photons calculated taking also the attenuation in the coaxial cables into account. The temperatures of the coaxial lines are assumed to be their high-temperature end. The values (in dB) of the attenuators on the other plates are given in the label. Choosing a value of 20\,dB provides a suficciently low enough amount of thermal photons to the cell (green circles).}
\end{figure}

The amount of attenuation needed is determined by how low one requires the population of thermal photons to be at the experiment and the distribution of the attenuation, as well as the choice of the microwave components that need to be carefully chosen for compatibility with the operation of an adiabatic nuclear demagnetisation refrigerator. For a typical qubit experiment a population of thermal photons at $6\,$GHz well below $10^{-3}$ is desired \cite{krinner2019}. In order to achieve that, one needs at least $60\,$dB attenuation on the input line. The most effective way to reduce the amount of thermal photons is to install a significant amount of the attenuation on the coldest plate of the fridge, but that also leads to excessive dissipation of heat. Therefore, the attenuators on the drive line have been distributed between the different stages (see Figure~\ref{fig:rf}). To thermalise the central conductor in the coax lines, $0\,$dB attenuators were used where no attenuation was needed; this includes all stages for the output line, and pulse tube stage 1 plate (PT1P) and ANDRP for the input line. Here we further choose to have 0 dB attenuation on the input line at the ANDRP to minimise the active heat load. In Figure~\ref{fig:attenuation} we show that with 20 dB on the MCP stage we are still able to achieve a thermal photon population at $6\,$GHz below $10^{-3}$.

\begin{figure}
\includegraphics[width=1\linewidth]{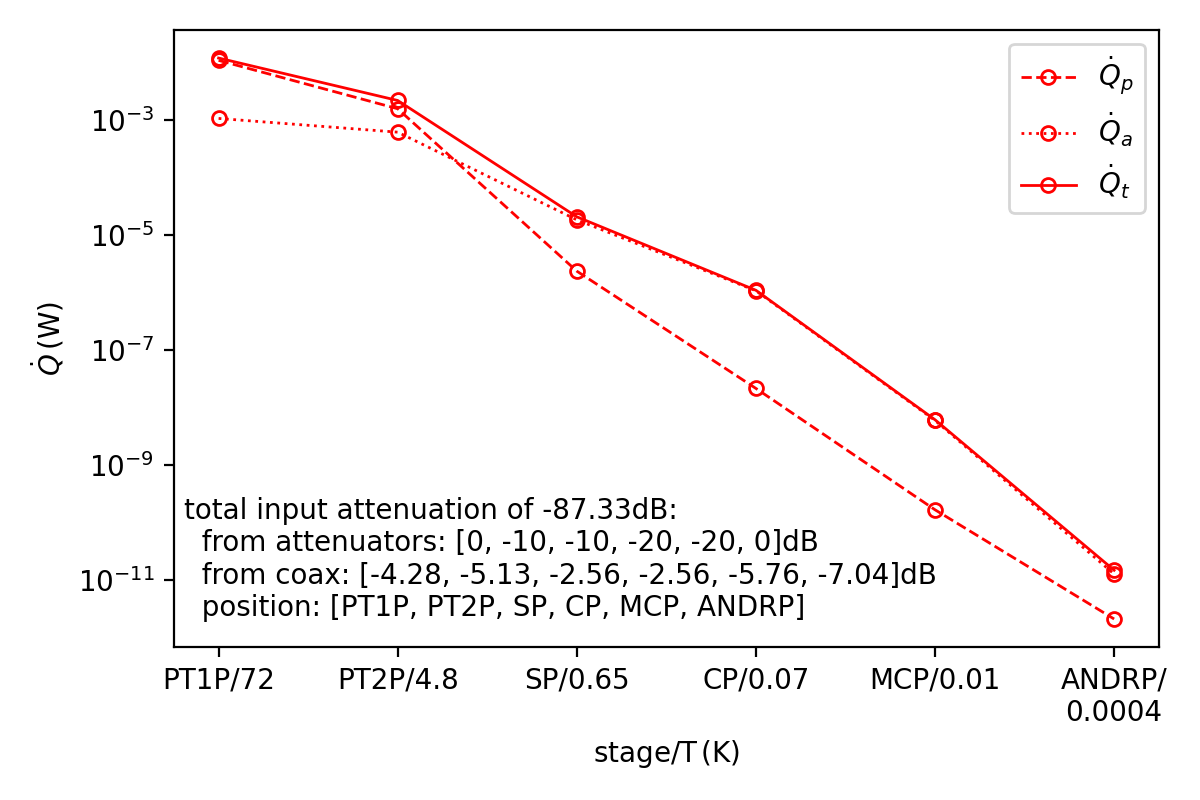}
\caption{\label{fig:dissipation}The passive and active loads coming from the RF-system to the different stages of the DR. The values used for the thermal conductivities were taken from \cite{pobell2007} and references therein, and the fit functions where extrapolated to the lower temperatures, hence overestimating the values for the passive loads. The description of how the attenuation in the cables is devided across the attenuators on the different stages is given in the main text. The active load is computed for an input signal power of $P_{\text{in}}=2.33\,$dBm, which provides $P_{\text{ cell}}=-85\,$dBm (the maximum operating power of the resonator) at the input of the cell.}
\end{figure}

The passive (due to thermal conductivity) and active (RF signals) thermal loads of the RF-setup on the stages of the DR are estimated with a simplified thermal model, which provides an upper-bound for the thermal load. In this model the attenuation in an input cable connecting two different stages of the DR, is added to the attenuator connected to its low-temperature end. The attenuation in the input cables connecting the MCP to the IR filter and the ANDRP attenuator to the cell are added to the attenuators on the MCP and ANDRP, respectively. This is because these two cables are at the same temperature as their respective stages. In the present model only the passive load is considered in the output line. This is justified by the low-power level of the output signal exiting the cell (below $\le-85\,$dBm up to PT1P and $\le-45\,$dBm after amplification at PT1P) and the negligible attenuation in the NbTi-NbTi cables. The combined loads on the CP and the MCP are less than $2\,$\textmu W and $10\,$nW, respectively. Figure~\ref{fig:dissipation} also shows a total load to the ANDRP lower than $20\,$pW, which represents $0.5\,$\% of the typical residual heat leak to the ANDRP.

\subsection{\label{sec:cell}The experimental cell}
\begin{figure}
\includegraphics[width=0.9\linewidth]{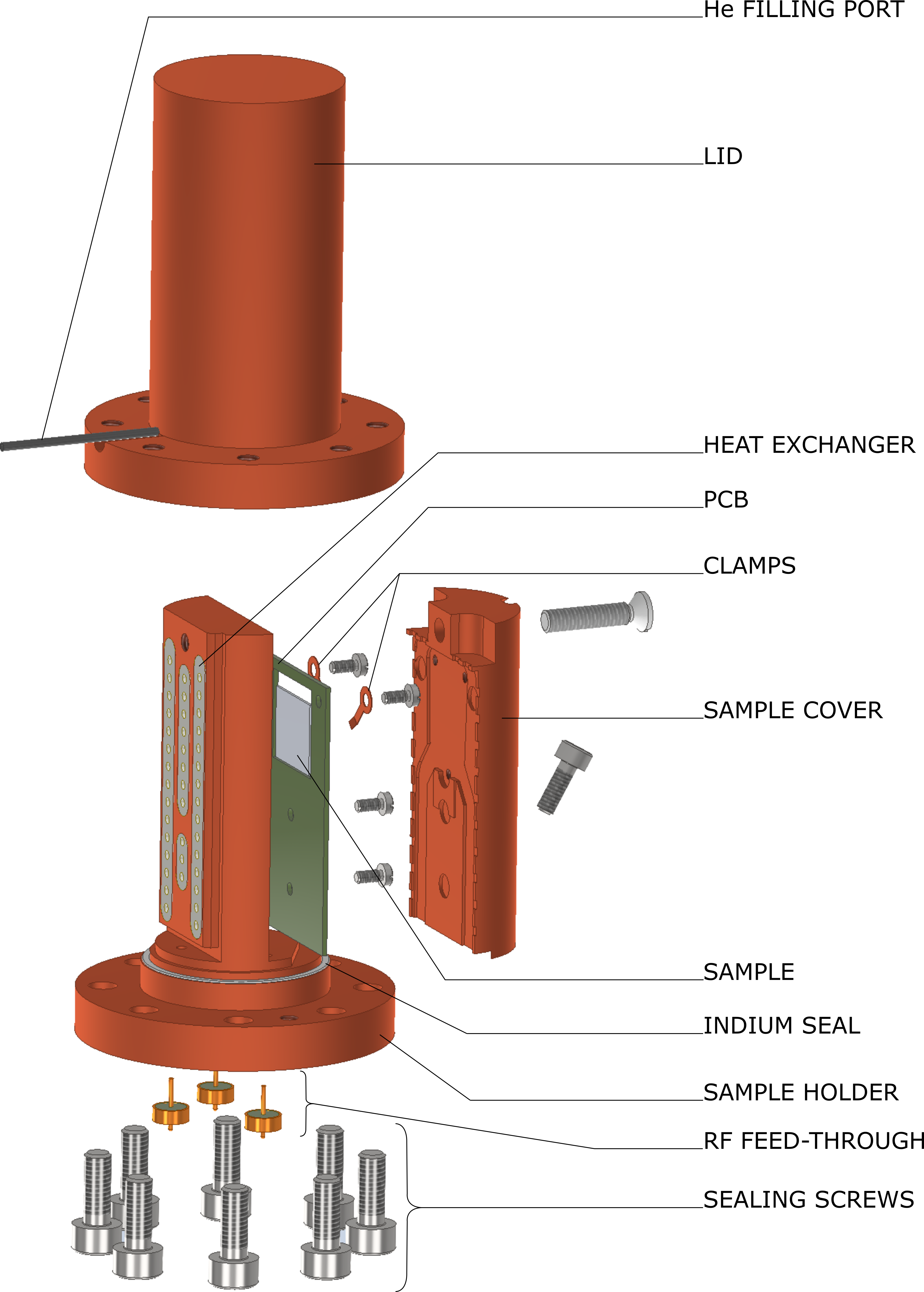}
\caption{\label{fig:cell}Exploded view of the He immersion cell used in the experiment. The lid, together with the base of the sample holder, the hermetic feed-through connectors and the indium seal, forms the leak tight enclosure of the immersion cell and is attached to the sample holder base with the eight sealing screws. A copper thermal link screwed in the base of the sample holder attaches and thermally connects the immersion cell to the ANDRP.}
\end{figure}%

Figure~\ref{fig:cell} shows an exploded view of the experimental cell, which is composed of three main elements: the sample holder, the sample cover and the lid. Each of these elements is itself composed of different parts, fulfilling specific functions. 

{\bf Sample holder}.
The body of the sample holder is made of oxygen-free high conductivity (OFHC) copper. The body has a step-shaped profile at its base for the indium seal and is directly thermalised to the ANDRP and provides thermal anchoring to all the other components of the cell.
The body is drilled through to receive three microwave feed-through connectors (one of which is shown in Figure~\ref{fig:cell}). These connectors are commercially available low-temperature hermetic feed-throughs. Two of the connectors are for the high-frequency in/out signals, and a third connector for a provisional DC-line. 

\begin{figure}
\includegraphics[width=0.9\linewidth]{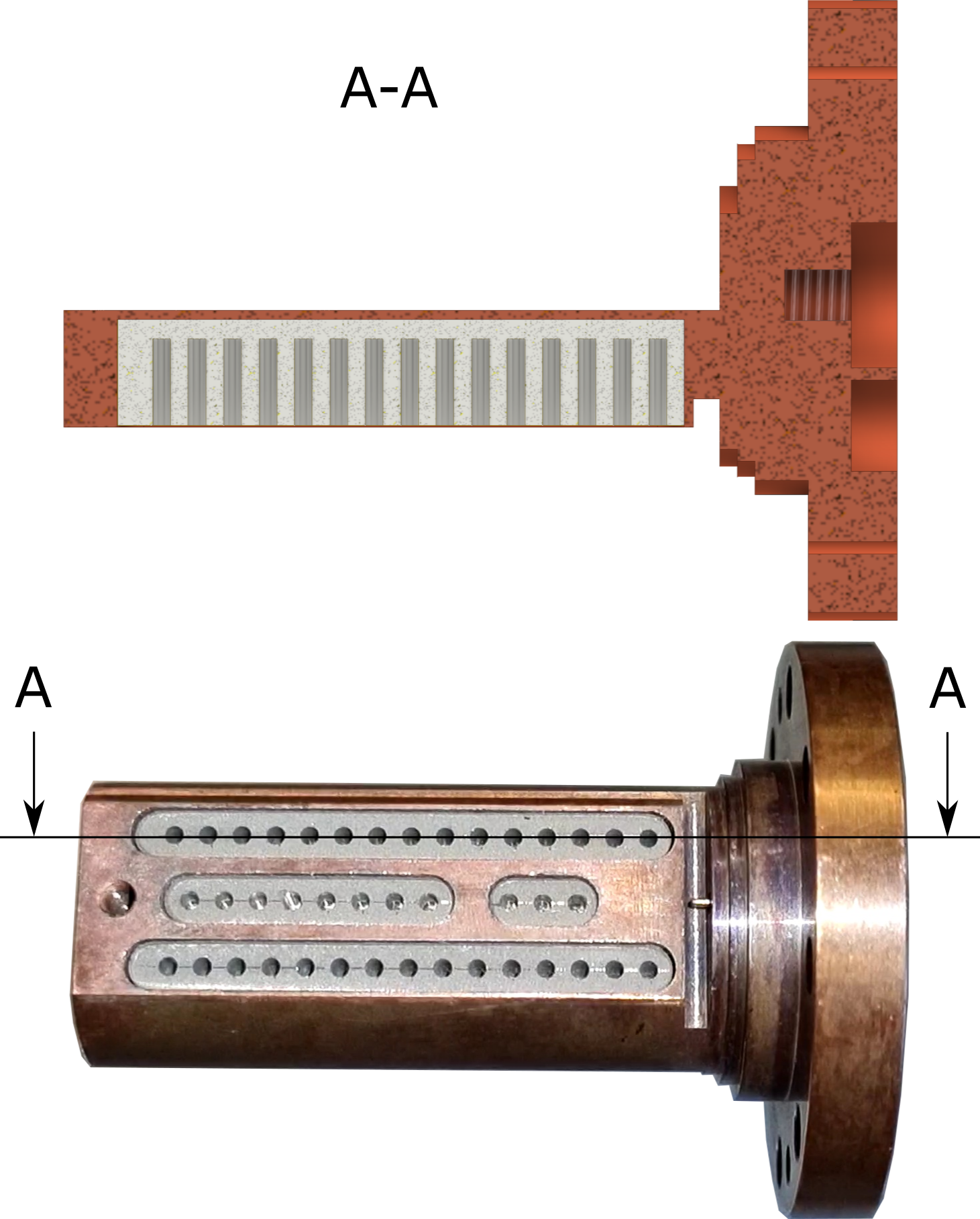}
\caption{\label{fig:sinter}Silver sinter heat exchanger used for the thermalisation of the liquid \textsuperscript{3}He, shown in grey on the CAD view at the top and on the photo at the bottom. The photo shows the back side of the sample holder with the four sinter pockets composing the heat exchanger of the immersion cell. Further details in the main text.}
\end{figure}

{\bf Heat exchanger}.
Because of the Kapitza boundary resistance, it is necessary to include heat exchanger in the immersion cell in order to cool down \textsuperscript{3}He below $1\,$mK. Although the microscopic mechanisms influencing the  the Kapitza resistance is not fully understood, there are well established recipes for how to improve thermal conductance of copper/liquid helium interfaces using silver sinter heat exchangers. Such a silver sinter heat exchanger has been made directly on the back of the sample holder, as shown in Figure~\ref{fig:sinter}. Holes have been drilled in the sinter to satisfy a rule of thumb, which states that that there should no be more than $1\,$mm of sinter between bulk liquid helium and bulk copper.  
A surface area of $24.1\,$m\textsuperscript{2} of the sinter has been measured with the BET (Brunauer–Emmett–Teller) N\textsubscript{2} isotherms method performed at $77\,$K. We use silver powder with a $70\,$nm grain size, giving an estimated Kapitza boundary resistance for this heat exchanger at the \textsuperscript{3}He/heat exchanger interface of \cite{pobell2007}
\begin{equation}
    R(T)_{\text{K}}=41.5\,T^{-1}\,\text{K/W.}
\end{equation}

During the experiment, the immersion cell was filled with $n_{^3\text{He}}=0.1\,$mol of liquid \textsuperscript{3}He, which has a heat capacity of
\begin{equation}
    C(T)_{^3\text{He}}=2.3\,T\,\text{J/K},
\end{equation}
up to $10\,$mK \cite{pobell2007}. This gives a time constant for cooling the \textsuperscript{3}He in the cell of 
\begin{equation}
    \tau=R_{\text{K}}C_{^3\text{He}}=95\,\text{s.}
\end{equation}

{\bf Sample cover}.
The sample cover is made of OFHC copper and has two main functions: it serves as a displacement volume to reduce the amount of bulk liquid of \hethree~inside the cell and it forms a RF-cavity around the sample. Minimising the amount of \hethree~required to fill the cell means that the dimensions of the cavity around the sample must be carefully designed to prevent interfering box resonance modes.

{\bf Lid and $^3$He fill line}.
The lid, made of OFHC copper, is attached to the base of the sample holder with eight stainless steel screws and an indium seal makes the cell leak tight. The total helium volume inside the cell is $3.98\,$cm\textsuperscript{3} and the volume is connected to the room-temperature \hethree~gas handling system via a $140\,$\textmu m inner-diameter ($0.5\,$mm outer-diameter) CuNi capillary. To condense and thermalise the \hethree~injected in the cell, at each stage of the DR the fill line is winded and silver-brazed around a $\frac{1}{4}"$ copper tube, which is clamped to the cryostat plate.

To prevent pressure changes in the cell that would result from variations in the temperature profile along the fill line, a $1\,$l ballast volume is connected to the fill line at room temperature. This ballast volume reduces the impact of the density changes of \hethree, which below pressures of $5\,$bar has a strong temperature dependence above $2\,$K \cite{dobbs_book}. Hence the density of the \hethree~is particularly sensitive to the temperature of the PT2P, which can fluctuate by up to $20\,$\% during normal operation for various reasons (ramping of the demagnetisation magnet, fluctuations of  the cooling water temperature in the compressor, etc.).
The gas contained in the ballast volume can be considered as ideal in constant volume and its pressure thus scales with its temperature. At the time of the experiment the typical variation of the room temperature over $24\,$hours was about $2\,^{\circ}$C around an average of $25\,^{\circ}$ C, which gives a typical pressure stability of $\approx0.7\,$\%. 

\subsection{Temperature of the helium bath}
There is no thermometer in our set up which directly measures the liquid \hethree~in the cell. Instead, we measure the temperature of the ANDRP with a CSNT. Because of non-zero thermal resistance, a temperature gradient appears along the thermal path between the \hethree~bath and the ANDRP. In order to give an upper bound on the temperature difference between the plate and the \hethree~in the steady state, we make the overestimated assumption that the sample (in operation) dissipates $\dot{Q}=50\,$pW directly into the \hethree~bath (this is 10 times more than what is expected as a result of any measurement) and that the sintered heat exchanger has the same temperature as the copper sample holder.

The temperature of the heat exchanger $T_{\text{hx}}$ can be obtained from the Wiedemann-Franz law:
\begin{equation}
    \dot{Q}=\tfrac{L_0}{2R}(T_{\text{hx}}^2-T_{\text{adnr}}^2),
\end{equation}
where $L_0=2.44\times10^{-8}\,$W$\,\Omega\,$K$^{-1}$ is the universal Lorentz number, $R=1\,$\textmu$\Omega$ is an upper bound of the electrical resistance along the thermal path from the heat exchanger to the ANDRP, and $T_{\text{adnr}}$ is the temperature of the ANDRP. For $T_{\text{andr}}=400\,$\textmu K, this gives
\begin{equation}
    T_{\text{hx}}=\sqrt{T_{\text{andr}}^2+\tfrac{2R\dot{Q}}{L_0}}
    =405\,\text{\textmu K}.
\end{equation}

The temperature step at the \hethree/heat exchanger interface is given by $\dot{Q}R_{\text{K}}(T)=\Delta T$, 
where $R_{\text{K}}(T)$ is the temperature-dependent Kapitza boundary resistance given above and $\Delta T=T_{^3\text{He}}-T_{\text{hx}}$ is the positive temperature difference between the \hethree~bath and the heat exchanger. Considering $R_{\text{K}}(T_{\text{hx}}=405\,\mu\text{K})=102500\,$KW$^{-1}$, we arrive at $T_{^3\text{He}}=410\,$\textmu K; the temperature difference between the ANDRP and the liquid \hethree~bath is less than $2.5\,$\% at the coldest temperature ($T_{\text{andr}}=400\,$\textmu K) and drops to less than $0.1\,$\% at $T_{\text{andr}}=1\,$mK.

As another example we take the case of $N\sim 300$ used in the noise measurements in the main manuscript which shows clear temperature dependence down to 1 mK. The circulating power in the resonator is
$P_{\rm{circ}}(N) = N\hbar\omega_0^2$, 
and the amount of power dissipated into the TLS bath is given by 
\begin{equation}
\dot{Q}_{\rm{diss}}(N) = N\frac{\pi \hbar \omega_0^2}{2Q_i(N)}.\label{eq:qdiss}
\end{equation}
 This gives a dissipation $\dot{Q}_{\rm{diss}}\approx3$ fW into the TLS bath, which is taken away by the \hethree. 3 fW is about $10^6$ times smaller than the total heat leak to the nuclear stage of the fridge, and 1000 times smaller than the maximum experimental power assumed in the above estimate of the temperature gradient.

\subsection{\label{sec3}The solenoid for electron spin resonance measurements}
The solenoid for the electron spin resonance (ESR) experiment was designed to be compatible with the design of the immersion cell and with a targeted field-to-current ratio around 200\,mT/A. The precise value of the latter has been deduced from the ESR signature of the free-electron spins and is 220(2)\,mT/A.
The solenoid was wound with a $107\,$\textmu m single filament NbTi superconducting wire in CuNi cladding around a copper base with inner diameter of $24\,$mm and a wall thickness of $0.6\,$mm. For practical reasons, the ESR solenoid was not operated in persisted mode, as the ESR field was adjusted every tens of seconds. In order to prevent noise from room temperature instruments and power supplies to affect the magnetic field, a very low-pass filter with a cut-off frequency at $-3\,$dB of $f_{\text{c}}\approx20$\,mHz was used.

\section{Measurements}
Here we provide additional information regarding the measurement setup and procedures. Two samples were measured, the first one containing two resonators (Resonator A and B) of design as described in \cite{PhysRevApplied.14.044040}. The second chip with Resonators C, D, E (of design \cite{dejan}) was measured in a consecutive cooldown in the same immersion cell. Samples were fabricated in the same way, using NbN on sapphire. Most notable difference between designs was that the second chip (C, D, E) had a in-plane interdigitated capacitor gap and superconductor strip width of 2 $\mu$m, compared to 1 $\mu$m for Resonators A and B. This is to somewhat reduce the coupling strength to individual TLS while still predominantly probing surface TLS. Though resonators A and B were frequency tunable, this functionality was not used in the present experiments. Both the magnitude of the noise and the single photon loss in the two resonator designs were roughly the same.

We measure the frequency fluctuations of the resonators using the well-established Pound-locking technique \cite{lindstrom2011}. This technique is based on a single phase-modulated RF signal being passed down the measurement line in the cryostat.
Importantly, the modulation frequency is chosen (here $1.3$ MHz) such that phase modulation sidebands are outside the resonator bandwidth.
When this signal passes through the sample 
and after detection in a square-law detector (diode), it can be shown that the contributions from the lower and upper sidebands cancel due to symmetry if the carrier signal frequency is on resonance with the resonator ($f=\nu_0$). However, if not on resonance there will exist a low frequency component after the detector at the phase modulation frequency. 
The measurement electronics are configured to detect and null this error signal by adjusting the carrier frequency in a feedback loop. An advantage of this method for cryogenic measurements is that both signal and reference are passed through the same RF cables, and instabilities due to e.g. temperature drift at room temperature and inside the cryostat, vibrations or other sources that can affect the signal propagation are suppressed.

\begin{figure}
\includegraphics[width=1\linewidth]{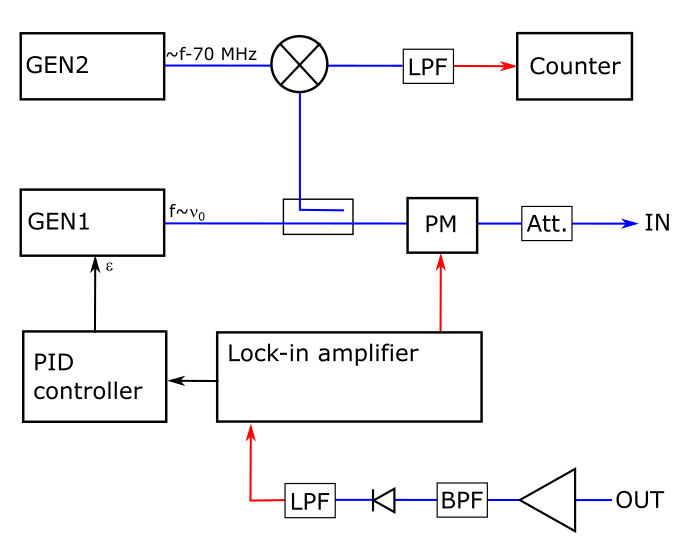}
\caption{\label{fig:poundsetup} Simplified schematic of the Pound measurement setup used to measure frequency noise. LPF is a low pass filter, BPF is a 40 MHz wide frequency tunable band pass filter, PM is a phase modulator (HMC538LP4), and Att. is a tunable attenuator (Vaunix LDA-5018V). Blue lines indicate high frequency (GHz) signals, red lines low frequency (MHz) and black very low frequency ($<$kHz) signals.}
\end{figure}

Figure \ref{fig:poundsetup} shows the schematics of the Pound-locking electronics used. As a stable frequency reference we use a SRS FS740 rubidium frequency standard, that also serves as the frequency counter to determine the instantaneous resonance frequency $\nu_0$. Using a gate time of $0.05$ s we record the gap-free time series of $\nu_0(t)$, that we analyse (see below) to extract the $1/f$ noise magnitude. A SMB100A RF generator is used to produce the carrier frequency that is modulated using an analogue phase modulator,  a $\sim 1$\,MHz phase modulation signal is supplied by a Zurich Instruments HF2LI lock-in amplifier. We tap off part of the signal from the carrier generator and downconvert it using a second generator (Hittite hmc-t2220) to about 70 MHz, within the bandwidth of the counter.
After amplification, filtering and detection in a RF diode the lock-in amplifier is used to demodulate the signal at the phase modulation frequency, and a PID controller (SRS SIM960) is given the task of nulling the lock-in signal by changing the frequency of the carrier generator operated in external frequency modulation mode.

\section{Noise analysis}

To measure the noise we record the resonator center frequency $\nu_0(t)$ with a gap-less sampling rate $\delta t = 0.05$ s. We transfer sampled blocks of 1000 data points from the frequency counter and for each such transfer we also record the ANDRP temperature, i.e. every 50 s. For long measurements versus temperature we then divide the recorded dataset into smaller chunks, each corresponding to some temperature interval $\langle T\rangle\pm \delta T$ where the error bar in temperature is taken as the maximum and minimum deviation from the mean in the interval.

The sampled $\nu_0(t)$ signal is converted to frequency noise spectral density $S_{\rm y}$ by calculating the overlapping Allan-variance $\sigma_{\rm y}^2(\tau)$ (AVAR) for M discrete samplings $\nu_k(n\tau)$ at multiples $n$ of the sampling rate $\tau$. 
\begin{equation}
\sigma_{\rm y}^2(n\tau) = \frac{1}{2(M-1)}\sum_{k=1}^{M-1}(\nu_{k+1}-\nu_k)^2.
\end{equation}
For $1/f$ noise the power spectral density $S_{\rm y}(f) = h_{-1}/f$ relates to the Allan variance as 
$\sigma_{\rm y}^2 = 2\ln{(2)}h_{-1}$, 
where $h_{-1}= A_0/2\pi$ \cite{rubiola}, where $A_0$ is the magnitude of the $1/f$ noise.
The AVAR is evaluated at several time-scales $t=n\tau$ ranging from $10^0$ to $10^{3}$ seconds and $h_{-1}$ is obtained by averaging the calculated AVAR across at least two decades in $\tau$. Error bars in $S_{\rm y}$ are calculated from the standard deviation of the AVAR in the same time interval. The selected range varies somewhat from measurement to measurement, however it is not changed within a measurement. The reason for this is that different resonators and different conditions may lead to other noise mechanisms (such as white noise) entering at short timescales, or drift at very long timescales. Occasionally very strong individual TLS appear and temporarily affect the AVAR data (temporal drift). Because of this we adapt the range of times for analysis in order to avoid over-estimating the noise. Admittedly, this approach works less well at low measurement powers, where strong TLS appear more frequently and on many different timescales during measurements.

\section{Extended data}

\subsection{Full noise data}
\begin{figure*}
\centering
\includegraphics[width=1\textwidth]{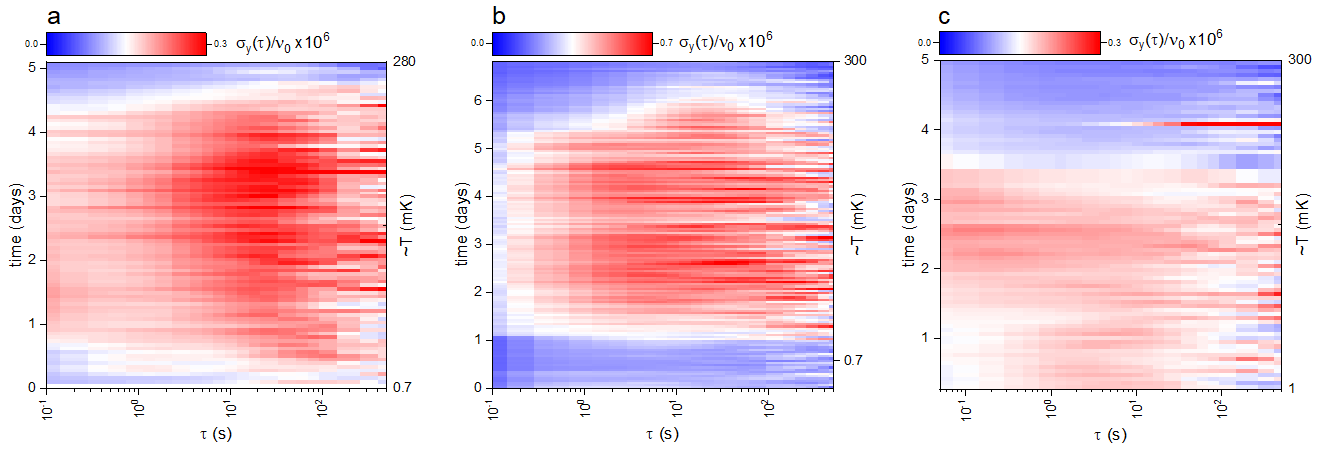}
\caption{\label{fig:full_sy} The calculated $\sigma_y(\tau)$ for the whole temperature ramp in Figure 3a in the main manuscript for a) Full cell $N\sim 400$, b) Full cell $N\sim20$ and c) Empty cell $N\sim 400$, highlighting the temporal drift of individual TLS at different timescales. Color scale is magnitude of the AVAR $\sigma_y(\tau)$ with blue corresponding to low values and red to higher values (note the different color scales in panels a-c). }
\end{figure*}

In figure \ref{fig:full_sy} we show the complete AVAR spectra behind the extracted data in Figure 3a. We plot the AVAR for each time-bin that corresponds to a specific temperature range as the temperature was ramped up. Rather than ambiguously trying to fit and subtract these from the overall flat background, or picking different background levels circumventing some of the random telegraph noise (RTN), we have instead included these in the analysis and calculation of the average noise level presented in Figure 3. In Figure 3 a large error bar is the result of a strong RTN fluctuator being present. 

In Figure \ref{fig:full_sy_chip2} we also show the measured temperature dependence of the noise from the second sample, resonator D, when the cell is filled with \hethree. The noise measurements take a significant amount of time ($\sim 6$ days per temperature ramp carried out at one drive power) and therefore we have only measured noise in some of the resonators. The noise in resonator D is following the same general trend as for resonators A and B. Because of the larger capacitor gap in resonator D compared to resonators A and B the magnitude of the noise is somewhat lower.

Table \ref{tab:noise_param} summarises the fitted temperature dependence of the noise from multiple measurements conducted, both in the high temperature regime (above $T_x$) and low temperature regime (below $T_x$).

\begin{figure}
\centering
\includegraphics[width=0.5\textwidth]{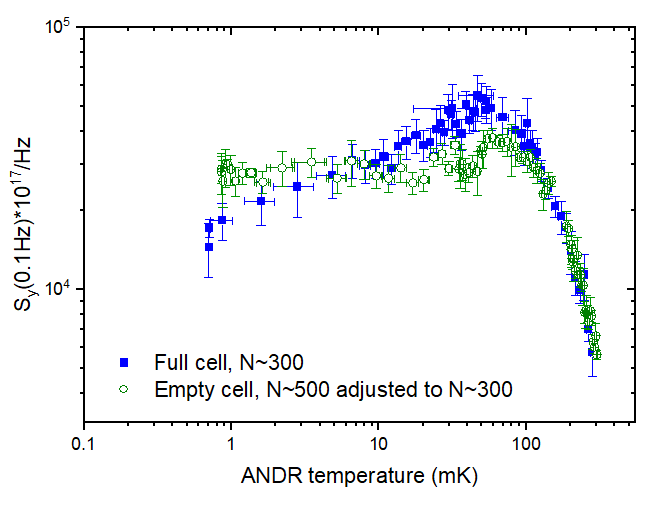}
\caption{\label{fig:syunscaled} The same data as in Figure 3a in the main manuscript without the 20x scaling of the noise for the empty cell data. Both measurements were conducted with the same input power to the sample, which results slightly different photon numbers because of the change in Q. The corresponding scaling factor 1.29 has been applied here to compare the data at the same photon numbers.}
\end{figure}

\begin{figure*}
\centering
\includegraphics[width=0.8\textwidth]{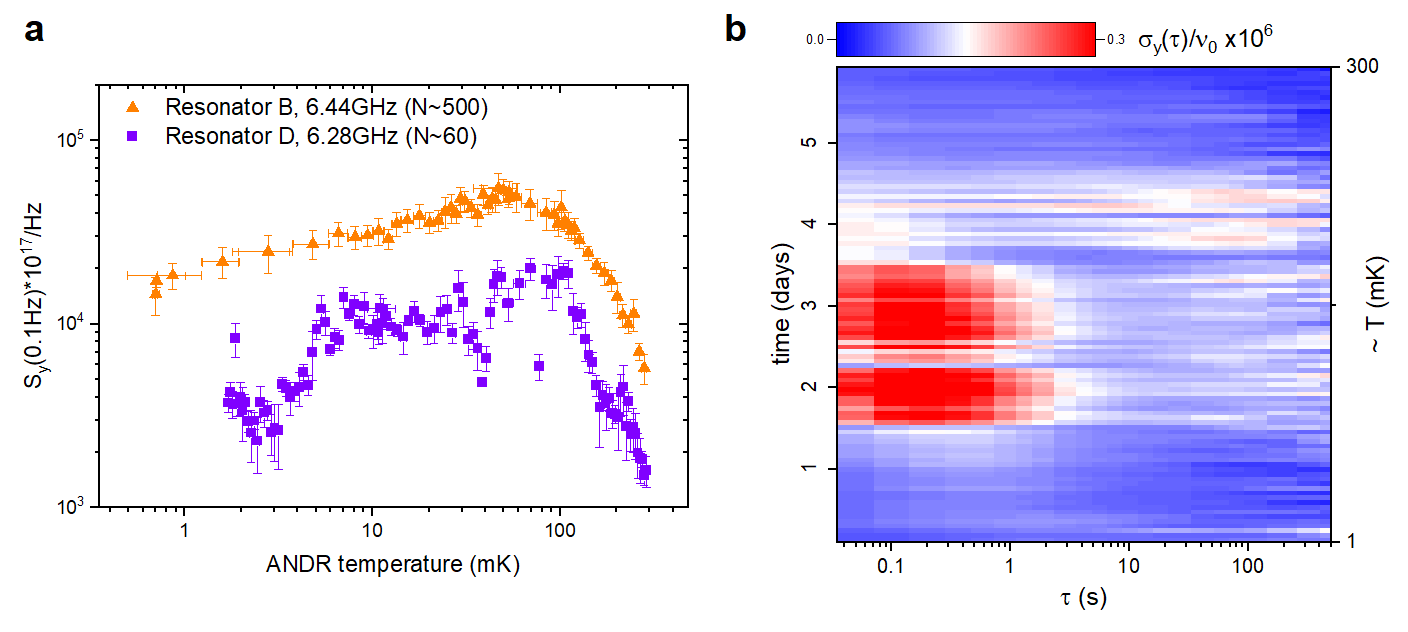}
\caption{\label{fig:full_sy_chip2} Noise in resonator D. a) the temperature dependence of the noise in the presence of \hethree. For reference we also compare with the data for resonator B also presented in Figure 3 of the manuscript. b) The calculated $\sigma_y(\tau)$ for the whole temperature ramp in a). Color scale is magnitude of the AVAR $\sigma_y(\tau)$ with blue corresponding to low values and red to higher values. }
\end{figure*}

In figure \ref{fig:syunscaled} we also present the same data as in Figure 3a of the manuscript, without the 20x scaling applied to the in-vacuum data presented in Figure 3a in the main manuscript. Here we only retain a small scaling factor of $\sqrt{N^{\rm empty}/N^{\rm full}}=1.29$ due to the small difference in photon number arising from the difference in $Q_i$ with full and empty cell (measurements were conducted with the same applied power to the sample).

\subsection{Power dependence of quality factor}
\begin{figure*}
\centering
\includegraphics[width=0.8\textwidth]{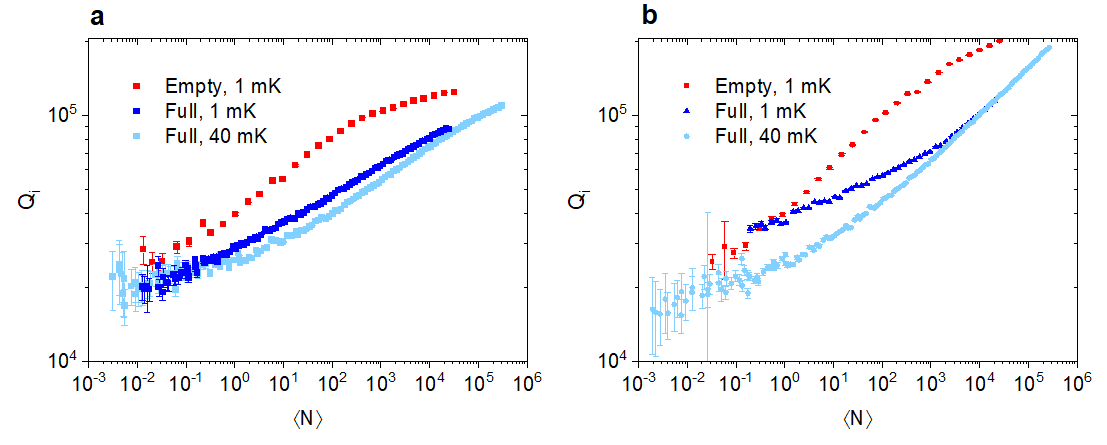}
\caption{\label{fig:qvsn} Measured internal quality factor of a) resonator A  (5.85 GHz) and b) resonator B (6.45 GHz) versus the average number of photons in the resonator. Resonator B was here subject to significant temporal fluctuations in the single photon $Q_i$ due to TLS.}
\end{figure*}

\begin{figure*}
\centering
\includegraphics[width=1\textwidth]{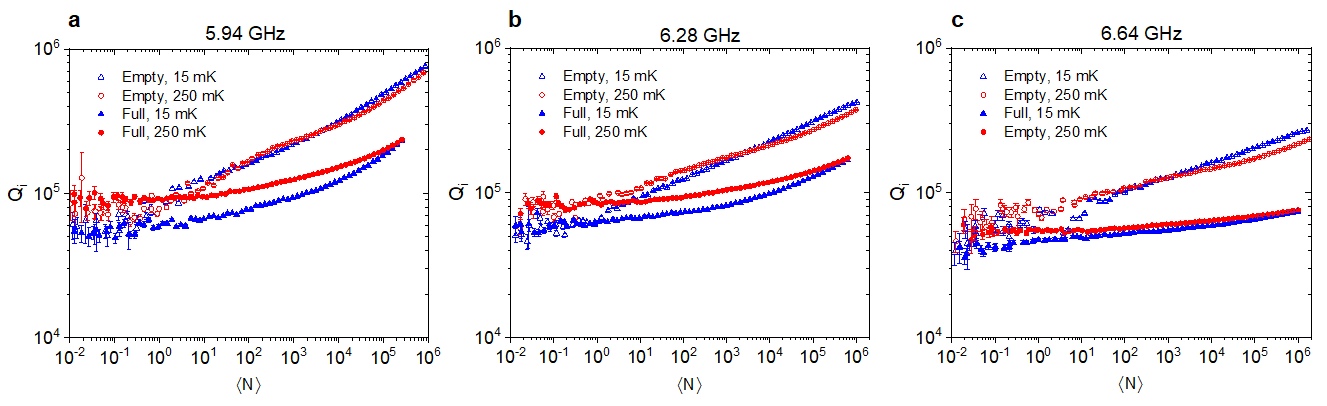}
\caption{\label{fig:QiCDF} Internal Quality factor versus average photon number for the three resonators C, D, and E, with and without \hethree~for two temperatures.}
\end{figure*}

Figures \ref{fig:qvsn} and \ref{fig:QiCDF} show the internal resonator quality factor versus the average photon number for different resonators on the two samples measured.
We note that a factor $\sim$1000 times enhancement in saturation power as a result of \hethree~filling was achieved on the second sample, and for the first sample the enhancement is somewhat smaller, 200-300 times. This can be due to a number of reasons, most likely due to the interface between \hethree~and TLS medium being not exactly the same. As the samples were fabricated at different times we may expect different ageing and surface contamination. It could also be related to possible presence of a (sub-)monolayer of \hefour~on the surface of resonators A and B. In the second cooldown (resonators, C, D and E) measures were taken to minimise \hefour~contamination further. We note that the different capacitor gap of the two samples is expected to contribute in the opposite way.

\subsection{Dielectric loss due to $^3$He}
\begin{figure*}
\centering
\includegraphics[width=0.7\textwidth]{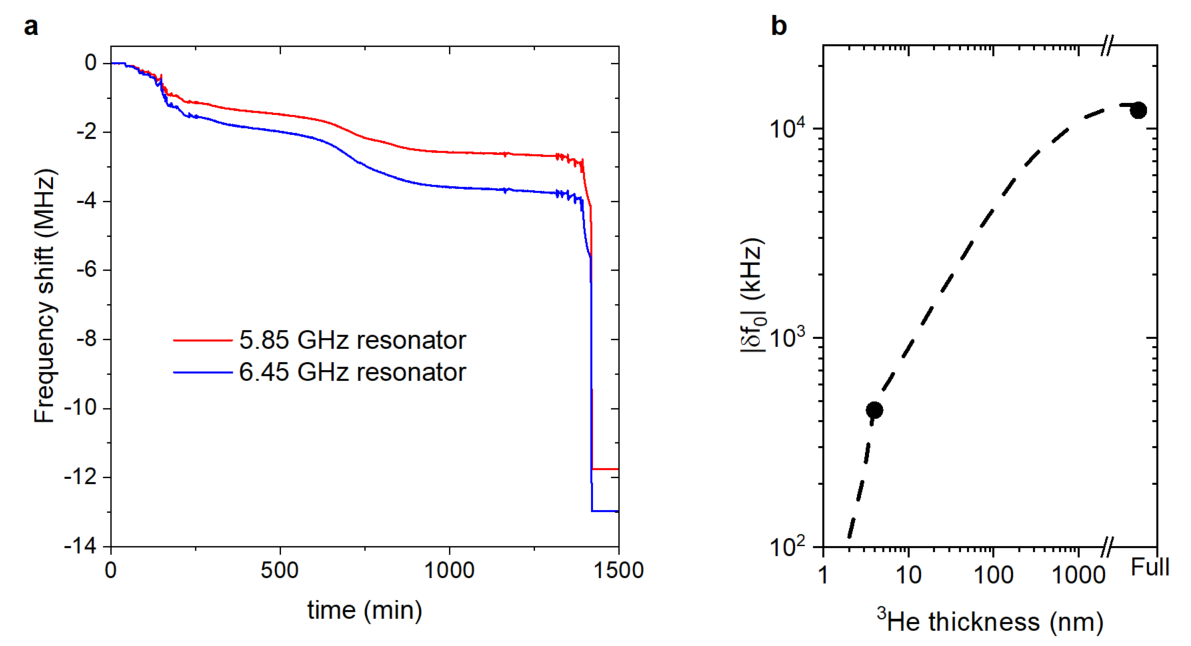}
\caption{\label{fig:3hethickness} Filling $^3$He. a) The change in center frequency of two resonators as the cell is filled with $^3$He. b) Expected frequency shift versus $^3$He film thickness. Markers are measured resonator frequency shift of the 5.85 GHz resonator in the presence of different amounts of \hethree~and the dashed line is the expected frequency shift obtained from COMSOL electrostatic simulations assuming a dielectric constant of \hethree~of 1.0426.}
\end{figure*}

The dielectric constant of \hethree~is known to be $\varepsilon_r = 1.0426$ at GHz frequencies and saturated vapor pressure. The dielectric constant is known to vary by a small amount with temperature and pressure of the liquid \hethree. In figure \ref{fig:3hethickness} we show the measured frequency shift of the 5.85 GHz resonator as a function of \hethree~thickness covering the sample. In one case the cell is completely filled, and in the other case it is covered with 10 monolayers of \hethree~(corresponding to $\sim4$\,nm), as estimated from the amount of \hethree~injected and the sinter surface area. We also show the expected frequency shift of the resonator assuming $\varepsilon_r = 1.0426$, calculated using COMSOL electrostatic simulations and assuming a substrate (sapphire) dielectric constant $\varepsilon_r = 10.4$. Frequency shift data is in agreement with simulation within one part in 1000. 

We estimate the \hethree~loss tangent from the single photon internal loss of our resonators. We fit the extracted TLS-limited internal Q to the empirical formula 
\begin{equation}
    Q_{\rm{int}}^{-1} = \frac{F\tan\delta}{(1+\langle N\rangle/n_c)^\alpha},\label{eq:qvsn}
\end{equation}
where $\langle N\rangle$ is the average photon number in the resonator, and $n_c$ is a critical photon number that depends on the TLS coupling strength to the resonator and their relaxation times. The standard tunneling model \cite{phillips} predicts an exponent $\alpha=0.5$ but in high Q superconducting resonators a value of $\alpha<0.5$ is often found, a signature of TLS-TLS interactions \cite{PhysRevLett.109.157005}. We summarise relevant extracted numbers in Table \ref{tab:res_param}.

\begin{table*}[]
    \centering
    \begin{tabular}{|c|c|c|c|}
    \hline
         & $\Delta t$ (days) &$f_0$ (MHz) & $Q_{i,n=1} \times 10^4 $ $(F\tan\delta)^{-1}$ \\
         \hline
        Resonator A, empty 330mK & 0 & 5839 & $3.7\pm 0.2 $ \\
        Resonator A, empty 5mK & 0 & 5839 & $3.2\pm 0.2 $ \\
        Resonator A, empty 1mK & 1 & 5839 & $3.4\pm 0.2 $ \\
        Resonator A, full 1mK & 13 & 5839 - 12.2& $2.5\pm 0.2 $ \\
        Resonator A, full 1mK & 44 & 5839 - 12.2& $2.0\pm 0.1 $ \\
        Resonator A, full 40mK & 45 & 5839 - 12.2& $2.2\pm 0.1 $ \\
        Resonator A, full 40mK & 47 & 5839 - 12.2& $2.3\pm 0.1 $ \\
        Resonator A, full 200mK & 59 & 5839 - 12.2& $3.0\pm 0.1 $ \\
        Resonator A, empty 15mK &66& 5839 & $2.8\pm 0.1 $ \\
        Resonator A, empty 15mK &67& 5839 & $2.8\pm 0.1 $ \\\hline
        Resonator B, empty 330mK& 0 & 6449& $3.9\pm 0.2 $ \\
        Resonator B, empty 5mK& 0 & 6449& $3.0\pm 0.3 $ \\
        Resonator B, empty 1mK& 1 & 6449& $2.9\pm 0.3 $ \\
        Resonator B, full 1mK& 13 & 6449 - 13.5& $3.7\pm 0.2 $\\
        Resonator B, full 1mK& 44 & 6449 - 13.5& $2.3\pm 0.3 $\\
        Resonator B, full 40mK& 45 & 6449 - 13.5& $2.1\pm 0.1 $\\
        Resonator B, full 40mK& 47 & 6449 - 13.5& $2.3\pm 0.1 $\\
        Resonator B, full 200mK& 59 & 6449 - 13.5& $3.0\pm 0.1 $\\
        Resonator B, empty 15mK& 66 & 6449 & $2.3\pm 0.1 $ \\
        Resonator B, empty 15mK& 67 &  6449 & $1.9\pm 0.1 $\\\hline
        Resonator C, empty 15mK &0 &5947.6 &$6.6\pm 0.8$ \\
        Resonator C, full 15mK &8& 5947.6-12.8 &$ 6.0\pm0.2$ \\
        Resonator C, full 15mK &8 &5947.6-12.8 &$7.4\pm0.2$ \\
        Resonator D, empty 15mK &0 &6285.6 &$5.4\pm0.3$\\
        Resonator D, full 15mK &8 &6285.6-13.5 &$6.4\pm0.2$ \\
        Resonator D, full 15mK &8 &6285.6-13.5 &$6.3\pm0.1$ \\
        Resonator E, empty 15mK &0 &6658.1 &$5.4\pm0.4$ \\
        Resonator E, full 15mK &8 &6658.1-14.1 &$4.1\pm0.2$ \\
        Resonator E, full 15mK &8 &6658.1-14.1 &$4.5\pm0.1$ \\\hline
    \end{tabular}
    \caption{Extracted resonator parameters with and without \hethree. Ranges given are the 95\% confidence bounds from fits to eq. (\ref{eq:qvsn}).}
    \label{tab:res_param}
\end{table*}

There is no noticeable effect of \hethree~on the single photon $Q_i$, instead the variations seen can be attributed to temporal drift due to TLS. Based on this we conclude that the loss introduced by \hethree~is much smaller than the fluctuations, and we take the mean of all observations and evaluate the upper limit for $F\tan\delta \ll (1/\langle Q_i^{\rm full}\rangle-1/\langle Q_i^{\rm empty}\rangle)^{-1} = 1.5\times10^{-6}$. We note that the error intervals reported are errors from the fits and do not capture the temporal variations in the parameters.

Next we estimate the filling factor which will allow us to put a bound on the loss tangent. Using COMSOL for electrostatic simulations of the electric field magnitude in the relevant dielectric volumes we obtain
\begin{equation}
F=\frac{\int_{V_{^3\rm{He}}}\varepsilon_r|E^2|dV}{\int_{V_{^3\rm{He}}}\varepsilon_r|E^2|dV + \int_{V_{s}}\varepsilon_{s}|E^2|dV}=0.10.
\end{equation}
Here $s$ denotes the substrate and we use $\varepsilon_{s}=10.4$ for the sapphire substrate and $\varepsilon_{r}=1.0426$ for \hethree. We thus arrive at $\tan\delta \ll 1.5\times10^{-5}$. 

For a qubit with relaxation limited by dielectric loss we have $T_1=\nu_{\rm qubit}F_{\rm qubit}\tan\delta $. If \hethree~fills the whole volume above the qubit and using a substrate with similar dielectric constant, then $F_{\rm qubit} \approx F$, and hence we find a lower bound imposed by the dielectric loss in \hethree~on qubit coherence of $T_1>110$ $\mu$s for $\nu_{\rm qubit} = 6$ GHz.

\begin{table*}[]
    \centering
    \begin{tabular}{|c|c|}
    \hline
          From dataset & -1-2$\mu$ for $T>T_{x}$ \\\hline
       Resonator A, Full  & $-1.50\pm0.9$\\
       Resonator B, Full  & $-1.76\pm0.18 $\\
       Resonator B, Empty  & $-1.47\pm0.12$\\
       Resonator B, Empty  & $-1.42\pm0.16$ \\&\\\hline
          From dataset & $T^\beta$ for $T<T_{x}$ \\\hline
       Resonator A, Full  & $0.25\pm0.04$\\
       Resonator B, Full  & $0.24\pm0.005$\\
       \hline
    \end{tabular}
    \caption{Summary of extracted values for the parameter $\mu$ from different resonators and parts of data. Low T refers to fits to the noise at temperatures $T<20$ mK in the saturated TLS regime, and high T to the region $90<T<250$ mK where the scaling $T^{-1-2\mu}$ is expected to hold, this range is avoiding the cross-over region and not exceeding $hf>k_bT$.}
    \label{tab:noise_param}
\end{table*}

\subsection{Cooling in the presence of a thin $^3$He film}
A frequency shift of 450 kHz was observed for resonator A as a small amount of \hethree~was first condensed in the sample cell at 300 mK. From the amount of \hethree~injected and the sinter surface area we estimate a thin film of approximately 10 atomic layers ($\sim 4$ nm) covering the surface of the sample. The 450 kHz shift is 3.7\% of the total frequency shift seen when fully immersed. This is in good agreement with electrostatic simulations of the resonator frequency shift assuming a dielectric constant of \hethree~of 1.0426, as we show in Figure \ref{fig:3hethickness}.

\begin{figure}
\centering
\includegraphics[width=0.5\textwidth]{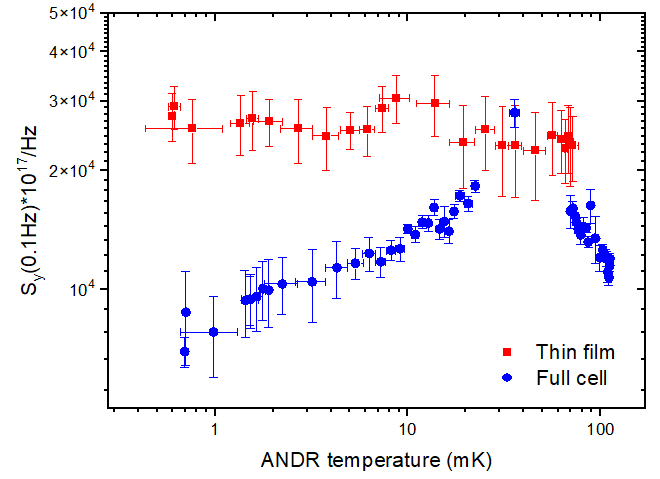}
\caption{\label{fig:some3he} Noise for resonator A (5.85 GHz) compared with cell filled completely with \hethree~and with just a thing layer on the surface. In both cases $\langle N\rangle \sim 400$. Gaps in the data for the full cell is due to a hardware fault during this time. }
\end{figure}

Figure \ref{fig:some3he} compares noise measured with a thin \hethree~film coverage against the noise in the completely filled cell. For the thin \hethree~film the magnitude of the noise is constant below $T\sim 100$ mK, the same as for the empty cell. The data for the filled cell is in good agreement for the trend observed also for resonator B (see main text). 

\begin{figure}
\centering
\includegraphics[width=0.5\textwidth]{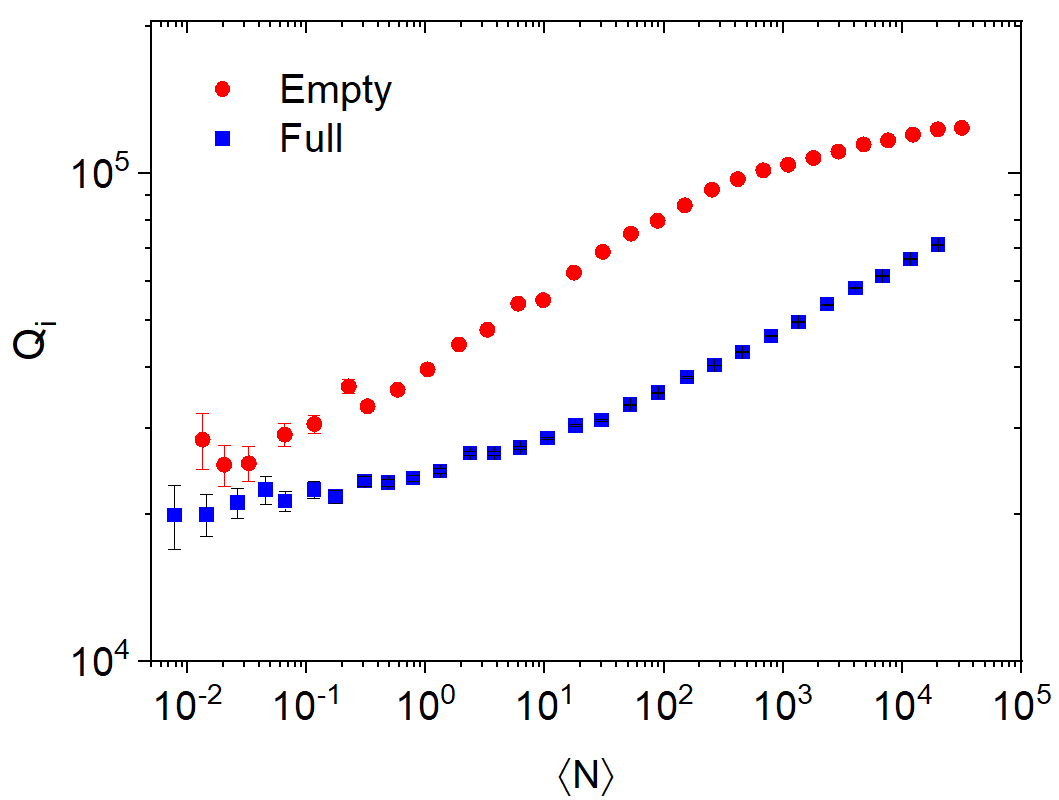}
\caption{\label{fig:some3he_q} Internal Quality factor versus average photon number for resonator A (5.85 GHz) in the empty (vacuum) cell and with just a thin layer of \hethree~on the surface. Measured with a ANDRP temperature of 1\,mK.}
\end{figure}

It thus appears that just a small amount of \hethree~does not cool down the TLS bath sufficiently to observe any clear change in the noise compared to vacuum. However, the drastic change in saturation power (measured from the power dependence of the resonator) is still observed even with a small amount of \hethree~present, shown in Figure \ref{fig:some3he_q}. This is consistent with the picture of \hethree~strongly interacting with the TLS, increasing their energy relaxation. To the contrary, with just a thin film of \hethree~the cooling of the \hethree~itself is expected to be very poor, and there may not even be a continuous thermal link to the sinter heat exchangers and the ANDRP.

\begin{figure}
\centering
\includegraphics[width=0.5\textwidth]{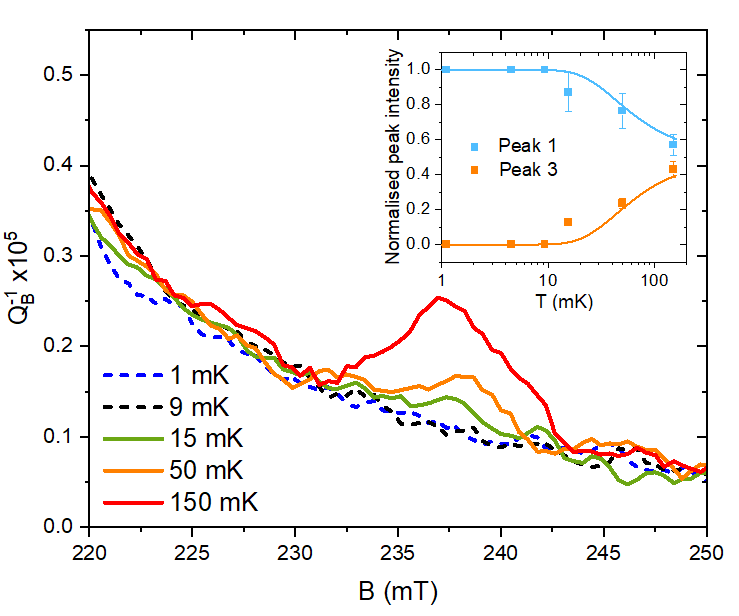}
\caption{\label{fig:esr_some3he} ESR peak intensity for the high-field atomic hydrogen peak versus nuclear stage temperature in the presence of a thin \hethree~film on the sample surface. $\langle N\rangle\sim 10^3$. The inset shows the fitted normalised peak intensity and the expected intensity due to thermal population of the ESR levels split by 1.42 GHz. At 1, 5 and 9 mK no trace of a peak is observed above the noise and datapoints are set to zero.}
\end{figure}

We also investigated the cooling of surface spins with the same small amount of \hethree~present. Figure \ref{fig:esr_some3he} shows the temperature dependence of the second hydrogen peak which is expected to have a population and peak intensity vanishing as temperature goes to zero. Here we find that even a small amount of \hethree~is enough to cool the surface spins.

\subsection{Pressure dependence}
\begin{figure*}
\centering
\includegraphics[width=0.6\textwidth]{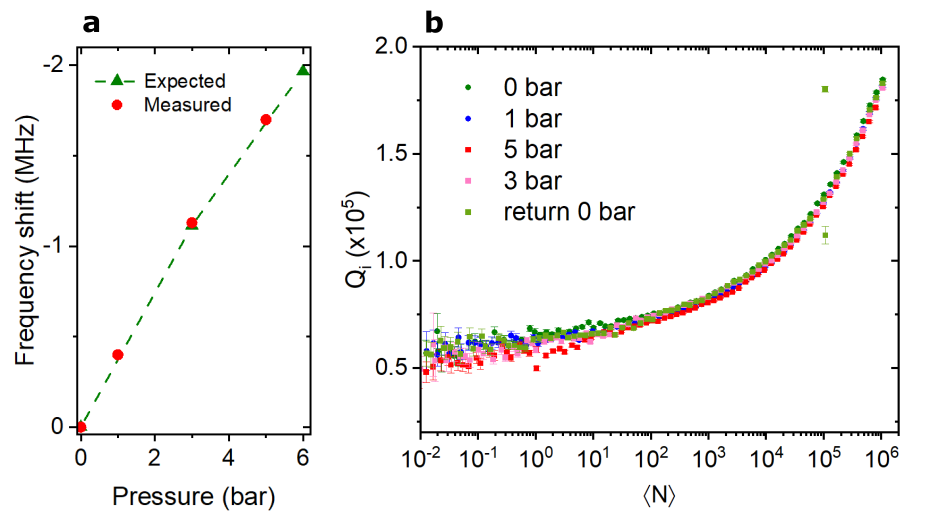}
\caption{\label{fig:pressure} a) Increasing the \hethree~pressure modifies the dielectric constant according to the Clausius-Mossotti relation. Using tabulated data for \hethree~density and speed of sound, and a zero pressure dielectric constant $\varepsilon_r = 1.0426$ we calculate the expected frequency shift as a function of applied pressure. The expected frequency shift is in good agreement the measured shift, thus confirming that the \hethree~pressure in the cell matches that at room temperature where the pressure gauge is located. Data obtained at 10 mK for the 6.26 GHz resonator. b) Internal Quality factor vs average photon number for the same resonator for pressures up to 5 bar. A small change my be present, but we cannot exclude the possibility that this is due to parameter fluctuations.}
\end{figure*}

Experiments at elevated pressures of \hethree~were conducted at 10 mK. We track the resonance frequency as we apply pressure, and plot the change in frequency in figure \ref{fig:pressure}a. From the measured dielectric constant and the known change in \hethree~density and speed of sound\cite{dobbs_book} we can calculate the expected frequency shift using the Clausius-Mossotti relation. This is found to be in excellent agreement with experimental data, confirming that the applied pressure is what we expect. In figure \ref{fig:pressure}b we then show the internal quality factor of the resonator versus average photon number, and compare it to the case of zero pressure. We find that increasing the pressure to 5 bar has a very small effect on the saturation power, increasing it about 20\%, a precise number is hard to estimate due to TLS parameter drift. Measurements at still higher pressure were not possible due to limitations of our immersion cell design.

\subsection{Additional data on noise at sub-mK temperatures}
\begin{figure*}
\centering
\includegraphics[width=0.8\textwidth]{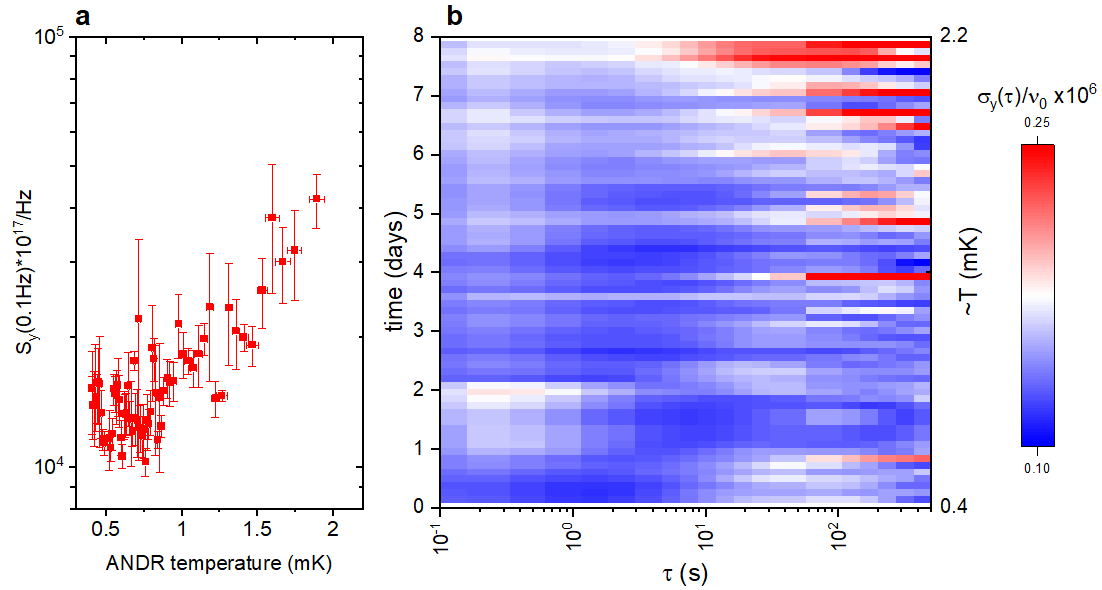}
\caption{\label{fig:full_sy_lowt} a) A slow warmup experiment exploring the lowest accessible temperatures. $\langle N\rangle\sim 400$. b) The calculated $\sigma_y(\tau)$ across all measured timescales, highlighting the temporal drift of individual TLS at different timescales.}
\end{figure*}

Figure \ref{fig:full_sy_lowt} shows a separate measurement where the sample was allowed to very slowly warm up from the lowest possible temperature of $\sim 400$ $\mu$K up to $\sim 2$ mK over the course of 8 days. Figure \ref{fig:full_sy_lowt}a shows the extracted magnitude of the 1/f noise and panel b shows the whole AVAR dataset. Even at these low temperatures TLS temporal fluctuations are evident as "bumps" appearing and disappearing on various timescales. 

While we cannot completely rule out TLS temporal drift, this additional data in Figure \ref{fig:full_sy_lowt} shows a clear increase of the noise starting at $T\sim 0.8$ mK. Below this temperature the noise remains constant with temperature, likely due to inefficient thermalisation. This also means that the low temperature increase in noise in the $N=20$ data presented in Figure 3a  in the main manuscript is due to TLS temporal drift. 
The $\sim 1$ mK saturation temperature is close to the superfluid transition of \hethree~occurring at $T_c=0.9$ mK at saturated vapour pressure. 
Hence the observed saturation at low T may suggest that the cooling of the low energy fluctuators becomes less efficient once \hethree~enters the superfluid state. However, we expect this transition to be smooth and the effect to be small near $T_c$, as the BCS gap in the \hethree~spectrum that opens up scales as $\Delta_{\rm^3He}=3.06k_bT_c(1-T/T_c)^{1/2}$.

\section{Theoretical models}

\subsection{Temperature dependence of the noise}
A detailed derivation of the theory for the generalised tunneling model (GTM) of interacting TLS at low temperatures can be found in \cite{PhysRevB.91.014201}. Here we give a brief summary of the GTM main result, the temperature dependence of the resonator frequency noise, which is valid in the regime $\Gamma_1\ll\Gamma_2$ for the TLS bath.  The model partitions the TLS in two categories: those that are high energy coherent fluctuators with energies $>k_BT$ and some of these are also (near-)resonant with the resonator. These couple to the resonator and absorb energy (which is then dissipated to phonons in the bulk via the TLS relaxation rate $\Gamma_1$). Thermal fluctuators are those TLS with an energy smaller than temeprature $E\ll k_BT$.
The random telegraph switching of these fluctuators contribute to the frequency noise if they strongly perturb a nearby coherent TLS, or they can contribute to the dephasing rate $\Gamma_2$ of the TLS if weakly coupled \cite{PhysRevB.91.014201}. 

The  low energy thermal fluctuators are assumed to have a poissonian probability distribution of switching rates $\gamma$ and a uniform energy distribution $\rho^F(E,\gamma) = \rho_0^F/\gamma$. A strongly coupled fluctuator is considered activated, i.e. contributing to the noise and undergoing random telegraph switching, if the temperature exceeds its energy scale. Hence the number of activated fluctuators scales as $\int_0^T\rho_0^FdE =  \rho_0^FT$.
The average number of activated fluctuators coupled to each resonant TLS is then given by $\mathcal{N}_F(T) = \frac{4\pi}{3}\rho_0^FR_0^3T$. Here $R_0$ is the interaction radius of the TLS, i.e. a fluctuator inside this radius couples strongly such as to shift the TLS energy more than its linewidth. 

We note that in \cite{PhysRevB.91.014201} $\mathcal{N}_F$ was estimated to be $\sim 1$ and this term dropped. Subsequent experiments \cite{burnett2016, degraaf2018, degraaf2021} revealed that in fact $\mathcal{N}_F\gg 1$ which means the magnitude of the $1/f$ noise acquires another temperature dependent prefactor, $\mathcal{N}_F$.

The frequency noise spectrum is defined as the autocorrelation of frequency fluctuations in the frequency domain
\begin{equation}
    \frac{S_{\delta\nu}}{\nu_0^2} = \lim_{\tau\rightarrow\infty}\frac{1}{\tau}\int_0^\tau\int_0^\tau\frac{\langle \delta\nu(t_1)\delta\nu(t_2)}{\nu_0^2}e^{i\omega(t_1-t_2)dt_1dt_2},
\end{equation}
 where $\nu_0$ is the resonance frequency.
Evaluating this in the limit of low temperature $k_bT\ll h\nu_0$ and including the contribution from multiple fluctuators coupled to each TLS gives

\begin{equation}
    \frac{S_{\delta\nu}}{\nu_0^2}(\omega) \sim \frac{8}{15}\langle d_0^4\rangle \frac{P_\gamma}{\omega} \frac{\chi}{U_0\Gamma_2}
    \mathcal{N}_F(T)F(\mathcal{E}),
\end{equation}
with 
\begin{equation}
    F(\mathcal{E}) = \left(\int_{V_h}dV\frac{|\mathcal{E}|^4}{\sqrt{1+|\mathcal{E}/\mathcal{E}_c|^2}}\right) / \left(\int_V\epsilon |\mathcal{E}|^2dV\right)^2.
\end{equation}

\noindent We further have that the TLS linewidth is given by 
\begin{equation}
   \Gamma_2 = c_0\chi\ln\left(\frac{\Gamma_1^{\rm max}}{\Gamma_1^{\rm min}}\right)\frac{T^{1+\mu}}{\nu_0^\mu},\label{eq:gtmG2}
\end{equation}
where $\Gamma_1^{\rm max}$ and $\Gamma_1^{\rm min}$ is the maximum and minimum phonon relaxation rates of the distribution of TLS in the bath \cite{PhysRevB.91.014201}.
This temperature dependence of the linewidth arises from the TLS interactions. Furthermore,
\begin{equation}
   \chi = P_0U_0\left(\frac{\nu_0}{E_{\rm max}}\right)^\mu\approx \tan\delta_i,
\end{equation}
which is independent of temperature and 
\begin{equation}
   \mathcal{E}_c = \frac{\sqrt{\Gamma_1\Gamma_2}}{2\langle d_0|\sin\theta|\rangle}
\end{equation}
is the critical electric field for saturation which inherits a temperature dependence only from $\Gamma_2$ such that $\mathcal{E}_c(T) \sim T^{(1+\mu)/2}$. $d_0$ is the TLS dipole moment and $\theta$ their relative angle to the local microwave electric field from the device. The critical photon number is proportional to the critical field squared: $N_c\propto\mathcal{E}_c^2(T)\propto \Gamma_1\Gamma_2$, which thus acquires a temperature dependence $T^{1+\mu}$.

\begin{figure}[t!]
\centering
\includegraphics[width=0.5\textwidth]{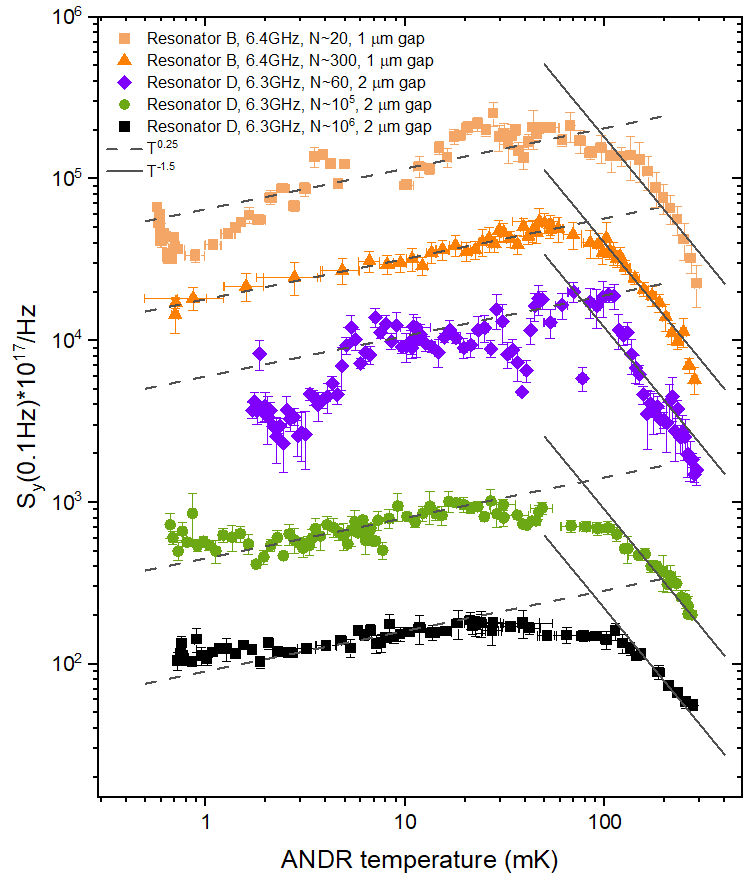}
\caption{\label{fig:tx} $1/f$ frequency noise as a function of temperature with the cell filled with \hethree. Comparison for several driving strengths ranging from $N\sim 60$ to $N\sim 10^6$ and for two different capacitor gap sizes in the resonators that otherwise have nominally the same frequency. The data shows that the crossover temperature $T_x$ remains the same under all these conditions and that the temperature dependence below $T_x$ appears universal. Several of the datasets the same as in other figures presented, here reproduced for convenience.}
\end{figure}

Taken together we can write the temperature dependent contributions to the resonator $1/f$ frequency noise as
\begin{equation}
    \frac{S_{\delta\nu}}{\nu_0^2}(\omega, T) \sim \frac{1}{\omega} \frac{T}{\Gamma_2}R_0^3
    F(\mathcal{E}).
\end{equation}
For weak driving fields, $\mathcal{E}/\mathcal{E}_c\ll1$ $F(\mathcal{E})$, becomes a temperature independent constant prefactor and TLS have a spectral width $\Gamma_2$, such that $R_0^3 = U_0/\Gamma_2$.  In the strong field regime the TLS are power broadened such that instead $R_0^3 = U_0/\Omega_R$, where $\Omega_R = 2\sin\theta d_0\cdot \mathcal{E}$ is the Rabi frequency and $U_0=d_0^2/\varepsilon$ is the dipole-dipole interaction strength. In the strong field regime we have also that $F(\mathcal{E})\propto \mathcal{E}_c$. Taken together 
\begin{equation}
    \frac{S_{\delta\nu}}{\nu_0^2}(\omega, T) \sim \begin{cases} 
    \frac{1}{\omega}\frac{T}{\Gamma_2^2}\propto T^{-1-2\mu} &{\rm for\hspace{3mm}} \mathcal{E}\ll \mathcal{E}_c\\
    \frac{1}{\omega}\frac{T\mathcal{E}_c}{\Gamma_2}\propto T^{(1-\mu)/2} & {\rm for\hspace{3mm}} \mathcal{E}\gg  \mathcal{E}_c.\label{eq:sy}\end{cases}
\end{equation}
Hence, for a constant driving power, starting at high temperatures and weak fields, we expect a crossover into the strong fields regime as we cool down the TLS bath and the TLS coherence increases, provided that $\Gamma_1\ll\Gamma_2$ at high temperature. With $\mu=0.25$ as we extract at high temperatures this would yield a low temperature power saturation scaling as $T^{0.375}$, somewhat different from what we observe.
It can also be understood that the crossover temperature $T_x$ acquires a power dependence $T_x\propto \langle N\rangle^{1/(2+2\mu)}$, such that if we increase the driving power, the crossover is expected to also move to higher temperatures. Likewise if the electric field strength in the device is diluted (by increasing the gap between metal electrodes in the resonator) $T_x$ should scale accordingly with electric field strength. In figure \ref{fig:tx} we compare several measurements varying the driving power by five orders of magnitude and and the capacitor gap by a factor 2, none of which has any effect on the crossover temperature, implying that TLS saturation cannot fully explain our results.

Contrary to eq. \ref{eq:sy}, if we instead are in the limit where relaxation dominates the TLS linewidth (for $T<T_x$) the GTM predicts
\begin{equation}
    \frac{S_{\delta\nu}}{\nu_0^2}(\omega, T)= \frac{1}{\omega}\frac{T}{\Gamma_1}R_0^3F(\mathcal{E})  \propto T,
\end{equation}
for all $\mathcal{E}$, assuming that $\Gamma_1$ is independent of temperature. I.e. in this regime the model does not match the observed $T^{0.25}$ trend.

Both in the relaxation limited regime and in the original GTM regime of $\Gamma_1\ll \Gamma_2$ we expect a power dependence of the noise according to $S_y\propto(1+\langle N\rangle/N_c)^{-1/2}$, as we also observe in Figure 3b. 

Finally we note that our conclusion that \hethree~only affects $\Gamma_1$ is consistent with the expected effect that \hethree~has on the TLS elastic phonon interaction energy $U_0^{ph}\approx M^2/\rho v^2$. Using the values for $M$, $\rho$, and $v$ for sapphire and \hethree~quoted in the main manuscript we estimate $U_0^{He}/U_0^{sap}\approx 0.2$. This means that dipolar interactions still dominate over elastic interactions in the presence of \hethree, in agreement with the fact that we observe no change in the interaction parameter $\mu$ nor the single photon $Q_i\propto \chi^{-1}\propto U_0^{-1}$.

\subsection{Power dependence of the quality factor}
Significant TLS interactions yield a logarithmic dependence of the quality factor on photon number\cite{PhysRevLett.109.157005, burnett2016}
\begin{equation}
   \frac{1}{Q_{\rm i}} = P_\gamma F \tan\delta \ln \left(c\sqrt{\frac{N_c}{\langle N\rangle}}\right).\label{eq:GTMq}
\end{equation}
Here $F\tan\delta$ is the single photon loss tangent, $P_\gamma$ is a constant of order one arising from the distribution of fluctuator switching rates, and $c$ is a constant quantified in \cite{PhysRevB.91.014201}. 
\begin{figure}[t!]
\centering
\includegraphics[width=0.45\textwidth]{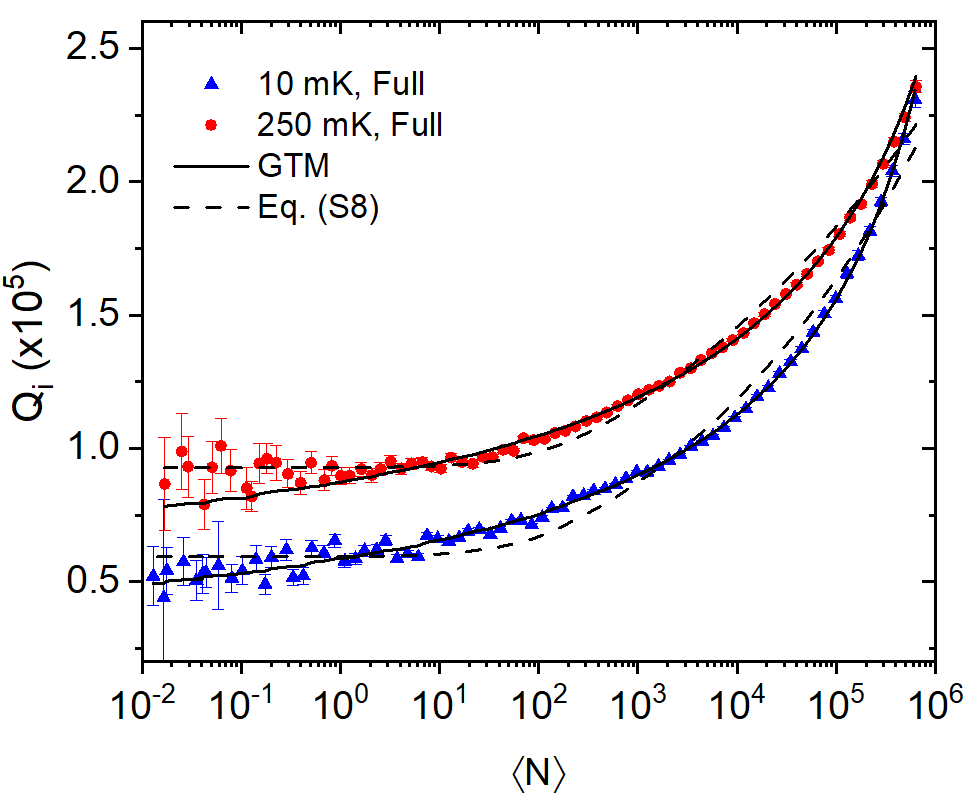}
\caption{\label{fig:Qistmvsgtm} Comparison of fits to eq. (\ref{eq:qvsn})) with $\alpha=0.11$, and GTM (eq. (\ref{eq:GTMq})) for two temperatures with the cell filled with \hethree.}
\end{figure}
Figure \ref{fig:Qistmvsgtm} compares typical $Q_i(\langle N \rangle)$ data with fits to the empirical formula Eq. (\ref{eq:qvsn}), and to the GTM expectation, Eq. (\ref{eq:GTMq}). There is exceptionally good agreement to the logarithmic GTM result, except at the very lowest photon numbers where noisy data obscures detailed comparison. 
Past comparison between these two models has often been challenging at mK temperatures, neither capturing exact dependencies precisely, which is still true for the vacuum measurements performed here. A reason for this is likely due to non-equilibrium effects and the interplay between overheating of the TLS bath and power saturation. In the presence of \hethree~we now achieve very good agreement, likely because the system is much closer to thermal equilibrium throughout the whole power range. We also note that for the evaluation of $F\tan\delta$ and the single photon loss values in Table \ref{tab:res_param} we still use Eq. (\ref{eq:qvsn}) for fits to remove the ambiguity from estimating the parameter $P_\gamma$.\\

\end{document}